\documentstyle[aps,multicol,psfig,eqsecnum]{revtex}  %...options for REVTEX 3.0 
\input epsf
\def\deltat{\tau}		% small parameter that controls the transition
\def\inte{g} 			% interaction strength
\def\numcop{{A}}
\def\disfac{\chi}
\def\dmhrs{{{\overline{\cal D}}^{\possym}} \Omega }
\def\dmwrs{{{\cal D}_{n}^{\possym}}\Omega }
\def \hzero { \hat 0}
\def \ofield#1{ \Omega(\hat #1)}
\def\HRS{{\rm HRS}}
\def\WRS{{\rm 1RS}}

\def\action{{\cal S}_{n}}
\def\mfnot{\overline}
\newcommand{\volfrac}{\varphi}	%... volume fraction

\newcommand{\oofield} [2] {\Omega({\hat #1}_{#2})}
\newcommand{\oopfield} [2] {\Omega^{#1}({\hat #2})}
\newcommand{\ooofield} [3] {\Omega^{#1}({\hat #2}_{#3})}

\def\possym{\dagger}
\def \qfield#1{ Q(\hat #1)}

\def\sumin{\sum_{j=1}^{N}}
\def\sumr{ \sum_{\alpha=0}^{n} }
\def \vzero { {\bf 0} }
\def\rmd{d}

\def\rmi{i}
\def\bnabla{\nabla}
\def\rbv{\hat{e}}
\def\wpeps{{\varepsilon}}
\def\bft{{\bf t}}
\def\nbft{{t}}
\newcommand{\eqbreak}{
\end{multicols}
\widetext
\noindent
\rule{.48\linewidth}{.1mm}\rule{.1mm}{.1cm}
}
\newcommand{\eqresume}{
\noindent
\rule{.52\linewidth}{.0mm}\rule[-.1cm]{.1mm}{.1cm}\rule{.48\linewidth}{.1mm}
\begin{multicols}{2}
\narrowtext
}
\begin{document} 
%----------------------------
%----------------------------
\draft 
\title{Renormalization-Group Approach 
to the Vulcanization Transition}
%\columnsep .375in 
%\twocolumn[ 
\author{Weiqun Peng and Paul M.~Goldbart} 
\address{Department of Physics, 
University of Illinois at Urbana-Champaign, \\
1110 West Green Street, 
Urbana, Illinois 61801-3080, U.S.A.}
%----------------------------
% \date{\today}		%...**
  \date{November 3, 1999}
%----------------------------
\maketitle
%----------------------------
\begin{abstract} 
%----------------------------
The vulcanization transition---the crosslink-density-controlled
equilibrium phase transition from the liquid to the amorphous solid
state---is explored analytically from a renormalization group
perspective.  The analysis centers on a minimal model which has
previously been shown to yield a rich and informative picture of
vulcanized matter at the mean-field level, including a connection with
mean-field percolation theory (i.e.~random graph theory).  This
minimal model accounts for both the thermal motion of the constituents
and the quenched random constraints imposed on their motion by the
crosslinks, as well as particle-particle repulsion which suppresses
density fluctuations and plays a pivotal role in determining the
symmetry structure (and hence properties) of the model.  A correlation
function involving fluctuations of the amorphous solid order
parameter, the behavior of which signals the vulcanization transition,
is examined, its physical meaning is elucidated, and the associated
susceptibility is constructed and analyzed.  A Ginzburg criterion
for the width (in crosslink density) of the critical region is
derived and is found to be consistent with a prediction due to de~Gennes.  
{\it Inter alia\/}, this criterion indicates that the upper critical 
dimension for the vulcanization transition is six.  Certain universal 
critical exponents characterizing the vulcanization transition are
computed, to lowest nontrivial order, within the framework of an
expansion around the upper critical dimension.  This expansion shows
that the connection between vulcanization and percolation extends
beyond mean-field theory, surviving the incorporation of fluctuations 
in the sense that pairs of physically analogous quantities (one 
percolation-related and one vulcanization-related) are found to be 
governed by identical critical exponents, at least to first order in 
the departure from the upper critical dimension (and presumably beyond).  
The relationship between the present approach to vulcanized matter and 
other approaches, such as those based on gelation/percolation ideas, is 
explored in the light of this connection.  To conclude, some expectations 
for how the vulcanization transition is realized in two dimensions, 
developed with H.~E.~Castillo, are discussed.
%----------------------------
\end{abstract}
%---------------------------- 
%----------------------------
%
% Possible Pacs numbers 
% (from 1999 Physics and Astronomy Classification Scheme (PACS)
% at http://www.aip.org/pacs/pacscheme.html) :
%
%       82.70.Gg  Gels and sols
%       61.43.-j  Disordered solids
%       64.60.Ak  Renormalization-group, fractal, and 
%               percolation studies of phase transitions 
%       61.43.Fs  Glasses
%
%----------------------------
% \begin{multicols}{2}
% \narrowtext
%----------------------------
%----------------------------
\noindent
\section{Introduction}
\label{SEC:Intro}
%----------------------------
Whilst a rather detailed description of the vulcanization transition
has emerged over the past few years within the context of a mean-field
approximation~\cite{prl_1987,PMGandAZprl,epl,cross}, the picture of
this transition beyond the mean-field level is less certain.  The
purpose of the present Paper is to provide a description of the
vulcanization transition beyond the mean-field approximation via the
application of renormalization group (RG) ideas to a model that
incorporates both the quenched randomness (central to systems
undergoing the vulcanization transition) and the thermal fluctuations
of the constituents (whose change in character is the fundamental
hallmark of the transition).  Our aim is to shed some light on certain
universal properties of the vulcanization transition within the
framework of the well-controlled and systematically improvable
approximation scheme that the RG provides, viz., an expansion about an
upper critical dimension that we shall see takes the value six.

We remind the reader that the vulcanization transition is an equilibrium 
phase transition from a liquid state of matter to an amorphous solid 
state.  (In addition to the technical reports cited 
above~\cite{prl_1987,PMGandAZprl,epl,cross}, we refer the reader to some 
informal accounts of the physics of the vulcanization transition~\cite{REF:SGRFchapZG,REF:TCrigidity,REF:PMGtrieste}.)\thinspace\ 
The transition occurs when a sufficient density of permanent random
constraints (e.g.~chemical crosslinks)---the quenched randomness---are
introduced to connect the constituents (e.g.~macromolecules), whose
locations are the thermally fluctuating variables.  In the resulting 
amorphous solid state, the thermal motion of (at least a fraction of) 
the constituents of the liquid undergo a qualitative change: no longer
wandering throughout the container, they are
instead localized in space at random positions about which they
execute thermal (i.e.~Brownian) motion characterized by random
r.m.s.~displacements.

Our approach to the vulcanization transition is based on a minimal
Landau-Wilson effective Hamiltonian that describes the energetics of
various order-parameter-field configurations, the order parameter in
question having been crafted to detect and diagnose amorphous
solidification.  This order parameter and effective Hamiltonian can be
derived (along with specific values for the coefficients of the terms
in the effective Hamiltonian) via the application of replica
statistical mechanics to a specific semi-microscopic model of randomly
crosslinked macromolecular systems (RCMSs), viz., the Deam-Edwards
model~\cite{REF:DeamEd}; this procedure is described in detail in
Ref.~\cite{cross}.  More generally, the form of the minimal model can 
be determined from the nature of the order parameter, especially its
transformation properties and certain symmetries that the effective
Hamiltonian need possess, along with the assumptions of the analyticity 
of the effective Hamiltonian and the continuity of the transition.  This 
system-nonspecific strategy for determining the minimal model was applied 
in Ref.~\cite{univ}.  There, it was shown that by regarding the effective
Hamiltonian as a Landau free energy one could recover from it the
mean-field description of both the liquid and emergent amorphous
solid states known earlier from the analysis of various
semi-microscopic models~\cite{epl,cross,REF:endlink,manifolds}.  
The mean-field value of order parameter in the solid state encodes a 
function rather than a number, and it possesses a certain mean-field
\lq\lq universality\rlap,\rq\rq\ by which we mean that (as the transition 
is approached from the amorphous solid side) both the exponent
governing the vanishing of the fraction of constituents localized
(i.e.~the gel fraction) and the scaled distribution of localization
lengths of the localized constituents turn out to depend not on the
coefficients in the Landau free energy but only on its qualitative
structure.  Support for this mean-field picture of the amorphous solid 
state, in the form of results for the localized fraction and scaled 
distribution of localization lengths, has emerged from extensive 
molecular dynamics computer simulations of three-dimensional, 
off-lattice, interacting, macromolecular systems, due to Barsky and
Plischke~\cite{REF:SJB_MP,REF:WherePub}.  In order to provide a
unified theory of the vulcanization transition that encompasses the
liquid, critical and random solid states, we shall in the present work
be adopting this Landau free energy as the appropriate Landau-Wilson
effective Hamiltonian.

We shall be focusing on the liquid and critical states, rather than 
the amorphous solid state, and shall therefore be concerned with the 
order-parameter correlator rather than its mean value.  Along the way, 
we shall therefore discuss the physical content of this correlator, why 
it signals the approaching amorphous solid state, and how it gives rise 
to an associated susceptibility whose divergence mark the vulcanization 
transition.

Given the apparent precision of the picture of the amorphous solid
state resulting from the mean-field
approximation~\cite{epl,cross,univ,REF:SJB_MP,REF:WherePub}, the reader 
may question the wisdom of our embarking on program that seeks to go beyond 
the mean-field approximation by incorporating the effects of fluctuations.  
We therefore now pause to explain what has motivated this program.
%----------------------------------------------------------------------
\hfil\break\noindent
(i)~Below six spatial dimensions, mean-field theory necessarily breaks
down sufficiently close to the vulcanization transition.  Although, as
we shall also see, the region of crosslink densities within which
fluctuations play an important role is narrower for dimensions closer
to (but below) six and for longer macromolecules, it is by no means
necessary for this region to be narrow for shorter macromolecules and
for lower-dimensional systems; thus, systems for which the
fluctuation-dominated regime is observably wide certainly exist.
%----------------------------------------------------------------------
\hfil\break\noindent
(ii)~Whilst there have been many successful treatments of critical
phenomena beyond the mean-field approximation in systems with quenched
randomness, these have, by and large, been for systems in which the
emergent order was not of the essentially random type under
consideration here or in the spin glass setting~\cite{REF:SpinGlass}.
Instead the emergent order has typically been of the type arising in
pure systems, albeit perturbed by the quenched disorder.  We are
motivated here by the challenge of going beyond mean-field theory in
the context of a transition to a structurally random state of matter.
%----------------------------------------------------------------------
\hfil\break\noindent
(iii)~The vulcanization transition has often been addressed from the
perspective of gelation/percolation
theories~\cite{REF:FloryBook,REF:Stauffer,REF:PGDGbook,REF:StCoAd,REF:LubIsaac}. 
Whilst this perspective can be (and certainly has been) taken beyond 
the mean-field level, it possesses but a {\it single\/} ensemble, and 
therefore does not incorporate the effects of {\it both\/} quenched 
randomness and thermal fluctuations~\cite{REF:singleENS}.  Given
that an essential aspect of the vulcanization transition is the impact
of the quenched random constraints on the thermal motion of the
constituents, the {\it a priori\/} identification of the vulcanization
transition with gelation/percolation is thus a nontrivial matter.  By
contrast with the gelation/percolation-type of approaches, the
analysis given in the present Paper applies directly to the
vulcanization transition exhibited by thermally fluctuating systems
and driven by quenched random constraints. It should therefore shed
some light on the relevance of the gelation/percolation-type
perspective for the vulcanization transition, as we shall discuss in
Sec.~\ref{SEC:Potts}.
%----------------------------------------------------------------------

This Paper is organized as follows.  
%----------------------------------------------------------------------
In Sec.~\ref{SEC:TechnicalFoundations} we give a brief account of 
the order parameter for the vulcanization transition, and of the 
Deam-Edwards replica approach to vulcanized matter and its 
field-theoretic representation, together with a minimal field-theoretic 
model for the vulcanization transition.  
%----------------------------------------------------------------------
In Sec.~\ref{SEC:MeanField} we summarize the mean-field--level picture
of the vulcanization transition, along with the picture of the amorphous 
solid state that emerges from it. 
%----------------------------------------------------------------------
In Sec.~\ref{SEC:Correlator} we discuss the order-parameter
correlator and susceptibility for the vulcanization transition, and
examine their physical content.
%----------------------------------------------------------------------
Having established this preparatory framework, we embark, in
Sec.~\ref{SEC:BeyondTree}, on the analysis of the vulcanization
transition beyond mean-field theory.  We begin by examining the
self-consistency of mean-field theory by estimating the impact of
fluctuations perturbatively, which results in the construction of a
Ginzburg criterion and the identification of six as being the
appropriate upper critical dimension. We then apply a momentum-shell
RG scheme to the minimal model, thus obtaining certain universal
critical exponents in an expansion around six dimensions.
%----------------------------------------------------------------------
Finally, in Sec.~\ref{SEC:Potts} we give some concluding remarks in
which we discuss connections between our approach and those based on
gelation/percolation, and we examine the role played by thermal
fluctuations, especially in lower spatial dimensionalities.
%----------------------------------------------------------------------
In three appendices we provide technical details associated with 
the derivation of the Ginzburg criterion,  we investigate the effects 
of various fields and vertices omitted from the minimal model, and we
present the full derivation of the RG flow equations. 
%----------------------------------------------------------------------
%------------------------------------------------------------------
%------------------------------------------------------------------
\section{Modeling the vulcanization transition\/}
\label{SEC:TechnicalFoundations}
%------------------------------------------------------------------
The purpose of the present section is to collect together the basic
ingredients of our approach to the vulcanization transition, including
the order parameter, underlying semi-microscopic model, replica field
theory, and minimal model.  All these elements have been discussed in
detail elsewhere, and we shall therefore be brief.  As the reader will
see, although its construction follows a quite conventional path, the
theory does possess some intricacies.  We shall therefore take various
opportunities to shed some light on the physical meaning of its
various ingredients.

Although most of our results are not specific to any particular system
undergoing a vulcanization transition, in order to make our presentation 
concrete we shall discuss the physical content for, and use notation 
specific to, the case of RCMSs.  We shall follow closely the notation
of Ref.~\cite{cross} and, accordingly, we shall adopt units of length
in which the characteristic size of the macromolecules is unity
(except in our discussion of the Ginzburg criterion,
Sec.~\ref{SEC:Ginzburg}).
%------------------------------------------------------------------
%------------------------------------------------------------------
\subsection{Order parameter for the vulcanization transition}
\label{SEC:OrderParameter}
%------------------------------------------------------------------
The appropriate order parameter for the vulcanization transition,
capable {\it inter alia\/} of distinguishing between the liquid and
amorphous solid states, is the following function of $\numcop$
wavevectors $\{{\bf k}^1,{\bf k}^2,\cdots,{\bf k}^{\numcop}\}$:
\begin{equation}
\Big[\,
\frac{1}{N} \sumin \int_{0}^{1}ds\,
\big\langle\exp i{\bf k}^{1}\cdot{\bf c}_{j}(s) \big\rangle_{\disfac}
\big\langle\exp i{\bf k}^{2}\cdot{\bf c}_{j}(s) \big\rangle_{\disfac}
\cdots
\big\langle\exp i{\bf k}^{\numcop}\cdot{\bf c}_{j}(s) \big\rangle_{\disfac}
\,\Big], 
\label{EQ:opDefinition}
\end{equation}
where $N$ is the total number of macromolecules, ${\bf c}_{j}(s)$
(with $j=1,\ldots, N$ and $0\leq s\leq 1$) is the position in
$d$-dimensional space of the monomer at fractional arclength $s$ along
the $j^{\rm th}$ macromolecule, $\langle\cdots\rangle_{\chi}$ denotes a
thermal average for a particular realization $\chi$ of the quenched
disorder (i.e.~the crosslinking), and $\left[\cdots\right]$ represents
a suitable averaging over this quenched disorder.  It is worth
emphasizing that the disorder resides in the specification of what
monomers are crosslinked together: the resulting constraints {\it do
not\/} explicitly break the translational symmetry of the system. In
the liquid state, for each monomer $(j, s)$ the thermal average
$\langle\exp i{\bf k}\cdot{\bf c}_{j}(s)\rangle_{\disfac}$ takes the
value $\delta^{(d)}_{{\bf k}, {\bf 0}}$ and thus the order parameter
is simply 
$\prod_{\alpha=1}^{\numcop}\delta^{(d)}_{{\bf k}^{\alpha},{\bf 0}}$.  
On the other hand, in the amorphous solid state we
expect a nonzero fraction of the monomers to be localized, and for
such monomers 
$\langle\exp i{\bf k}\cdot{\bf c}_{j}(s)\rangle_{\disfac}$ 
takes the form 
$\wp_{(j,s)}({\bf k})\,\exp i{\bf k}\cdot{\bf b}_{j}(s)$, 
i.e., a random phase-factor determined by
the random mean position ${\bf b}_{j}(s)$ of the monomer $(j,s)$ times
a random Debye-Waller factor $\wp_{(j,s)}({\bf k})$ describing the 
random extent to which the monomer is localized.  As reviewed in
Sec.~3 of Ref.~\cite{cross}, by choosing the wavevectors 
$\{{\bf k}^{\alpha}\}_{\alpha=1}^{\numcop}$ 
to satisfy the constraint 
${\bf k}^1 + {\bf k}^2 + \cdots + {\bf k}^{\numcop} = {\bf 0}$ 
the random phase-factors are
eliminated from the order parameter~(\ref{EQ:opDefinition}), and 
hence the order parameter is capable of distinguishing between the 
liquid and amorphous solid states and, furthermore, characterizing 
the randomness of the localization through its dependence on the 
collection of wavevectors.
%------------------------------------------------------------------
%------------------------------------------------------------------
\subsection{Replicated semi-microscopic model 
of vulcanized macromolecular systems}
\label{SEC:ReplicatedModel}
%------------------------------------------------------------------
Following Deam and Edwards~\cite{REF:DeamEd}, by (i)~starting from a
semi-microscopic Hamiltonian describing a system of macromolecules
interacting via an excluded-volume interaction, (ii)~introducing the
random constraints imposed by crosslinking, and (iii)~averaging over
the quenched disorder using the replica technique (with a physical
choice for the distribution of the disorder which leads to an
additional replica), one arrives at the disorder-averaged, replicated
partition function (for details, see Sec.~4 of Ref.~\cite{cross})
\begin{mathletters}
\begin{eqnarray}
[Z^n]&\propto&
\big\langle\exp(-{\cal H}_{n+1}^{\rm P})
\big\rangle_{n+1}^{\rm W}\,\,, 
\\
{\cal H}_{n+1}^{\rm P} 
&\equiv&
\frac{\lambda^{2}}{2}\sum_{j,j^{\prime}=1}^{N}
	\int_{0}^{1}ds
	\int_{0}^{1}ds^{\prime}\,
	\sum_{\alpha=0}^n
	\delta^{(d)}
	\big(
 		{\bf c}_{j}^\alpha  (s)
		-{\bf c}_{j^{\prime}}^\alpha (s^{\prime})
	\big)
% \nonumber\\&&\qquad\qquad
-\frac{\mu^{2}V}{2N}
	\sum_{j,j^{\prime}=1}^{N}
	\int\nolimits_{0}^{1}ds
	\int\nolimits_{0}^{1}ds^{\prime}\,
	\prod_{\alpha=0}^{n}
	\delta^{(d)}
	\big(
 		{\bf c}_{j}^{\alpha}(s)
		-{\bf c}_{j^{\prime}}^{\alpha}(s^{\prime})
	\big). 
\label{EQ:PureHam}
\end{eqnarray}% 
\end{mathletters}%
Here, $\langle\cdots\rangle_{n+1}^{\rm W}$ denotes a thermal average
taken with respect to the Hamiltonian for $n+1$ replicas of the
noninteracting, uncrosslinked system of macromolecules. Moreover,
${\cal H}_{n+1}^{\rm P}$ is an effective pure Hamiltonian accounting
for the interactions amongst the macromolecules and the effects of
crosslinking, the latter generating interactions between the replicas.
The parameter $\lambda^2$ measures the strength of the excluded-volume
interaction; the parameter $\mu^2$ measures the density of the
constraints and serves as the control parameter for the vulcanization
transition.  As a result of there being random {\it constraints\/}
rather than {\it interactions\/}, the coupling between the replicas
takes the form of {\it product\/} over all replicas rather than, say,
a {\it pairwise sum\/}.  As usual, the disorder-averaged free energy
is proportional to $[\ln Z]$, which we obtain via the replica
technique as $\lim_{n \rightarrow 0} {n}^{-1}\ln\big[Z^n\big]$.  Let
us mention, in passing, the symmetry content of this replica theory:
${\cal H}_{n+1}^{\rm P}$ is invariant under arbitrary independent
translations and rotations of the replicas as well as their
arbitrary permutation.

The natural collective coordinates for the vulcanization transition are
\begin{equation}
\qfield{k}
\equiv
\frac{1}{N}\sum_{j=1}^{N}
\int_{0}^{1}ds\,
\exp i\hat{k}\cdot\hat{c}_{j}(s), 
\label{EQ:DEF_Q}
\end{equation}
which emerge upon introducing Fourier representations of the two types 
of delta function in Eq.~(\ref{EQ:PureHam}), as discussed in detail in 
see Sec.~5.1 of Ref.~\cite{cross}.  (Such collective coordinates were 
first introduced in the context of crosslinked macromolecular melts by 
Ball and Edwards~\cite{REF:RCBallPaper}.)\thinspace\  
We use the symbol ${\hat k}$ to denote the replicated wavevector 
$\{{\bf k}^0, {\bf k}^1,\ldots, {\bf k}^n\}$, and define the 
extended scalar product $\hat{k}\cdot\hat{c}$ by 
${\bf k}^0\cdot{\bf c}^0+{\bf k}^1\cdot{\bf c}^1
+\cdots+{\bf k}^n\cdot{\bf c}^n$.  
The collective coordinates $\qfield{k}$ are the microscopic prototype of 
the order parameter~(\ref{EQ:opDefinition}), the latter being related to 
$\qfield{k}$ via $\lim_{n \rightarrow 0}
\big\langle\qfield{k}\big\rangle_{n+1}^{\rm P}$, 
where 
\begin{equation}
\big\langle \cdots \big\rangle_{n+1}^{\rm P}
\equiv
\frac{\big\langle\cdots
\exp(-{\cal H}_{n+1}^{\rm P})
\big\rangle_{n+1}^{\rm W}}
{\big\langle
\exp(-{\cal H}_{n+1}^{\rm P})
\big\rangle_{n+1}^{\rm W}}\,\,.
\end{equation}
%------------------------------------------------------------------
%------------------------------------------------------------------
\subsection{Replica field theory for vulcanized macromolecular systems}
\label{SEC:FieldTheory}
%------------------------------------------------------------------
As discussed in detail in Sec.~5.3 of Ref.~\cite{cross}, one can put
the partition function into a form of a field theory by applying a
Hubbard-Stratonovich transformation to the collective coordinates
$\qfield{k}$; we denote the corresponding auxiliary order-parameter
field by $\ofield{k}$.  At this stage one encounters a vital issue,
viz., that it is {\it essential\/} to draw the distinction between
examples of $\qfield{k}$ and $\ofield{k}$ that belong to the {\it
one-replica sector\/} (\WRS) and those that belong to the {\it
higher-replica sector\/} (\HRS).  The distinction lies in the value of
$\hat k$: for a replicated wavevector $\hat k$, if there is exactly
one replica for which the corresponding $d$-vector ${\bf k}^\alpha$ is
nonzero [e.g.~${\hat k}=(\vzero,\ldots,\vzero,{\bf
k}^\alpha\neq\vzero,
\vzero,\ldots,\vzero)$] then we say that $\hat k$ lies in the 
one-replica-sector ($\hat k \in \WRS$) and that the corresponding
$\qfield{k}$ and $\ofield{k}$ are \WRS\ quantities.  On the other
hand, if there is more than one replica for which the corresponding
components of ${\hat k}$ are nonzero then we say that $\hat k$ lies in
the higher-replica-sector ($\hat k \in \HRS$) and that the
corresponding $\qfield{k}$ and $\ofield{k}$ are \HRS\ quantities.  For
example, if ${\hat k}=(\vzero,\ldots,{\bf k}^\alpha\neq\vzero,\ldots,
{\bf k}^\beta\neq\vzero,\ldots,\vzero)$ then $\hat k$ lies in the
\HRS.  (More specifically, in this example $\hat k$ lies in the
two-replica sector of the \HRS.)\thinspace\ The importance of this
distinction between the \WRS\ and the \HRS\ lies in the fact, evident
from the order parameter~(\ref{EQ:opDefinition}), that the
vulcanization transition is detected by fields residing in the \HRS,
whereas the \WRS\ fields measure the local monomer density, and
neither exhibit critical fluctuations near the vulcanization
transition nor acquire a nonzero expectation value in the amorphous
solid state.

Bearing in mind this distinction between the \WRS\ and \HRS\ fields,
the aforementioned Hubbard-Stratonovich transformation leads to the
following field-theoretic representation of the disordered-averaged
replicated partition function:
\begin{equation}
[Z^{n}] 
\propto
\int\dmwrs 
\int\dmhrs\,
\exp\Big(-ndN{\cal F}_{n}
\big(\{
	\Omega^{\alpha}({\bf k}),
	\Omega({\hat k})
     \}\big)\Big),  
\label{EQ:RCMSpartition}
\end{equation}
where $\Omega^{\alpha}({\bf k})$ 
[which represents $\Omega({\hat k})$ when 
${\hat k}=(\vzero,\ldots,\vzero,
{\bf k}^\alpha={\bf k}\neq\vzero, 
\vzero,\ldots,\vzero)$] is a \WRS\ field, 
$\Omega({\hat k})$ is a \HRS\ field, and 
the explicit expressions for
the resulting effective Hamiltonian ${\cal F}_{n}$ and functional
integration measures~\cite{REF:Measure} are given by Eqs.~(5.12) and
(5.9) of Ref.~\cite{cross}. In this formulation of the statistical
mechanics of RCMSs, one can readily establish exact relationships
connecting average values and correlators of $\qfield{k}$ with those 
of $\ofield{k}$~\cite{REF:WasFoot}. (Such relationships between
expectation values involving microscopic variables and auxiliary
fields are common in the setting of field theories derived via
Hubbard-Stratonovich transformations~\cite{REF:Zinn}.)\thinspace\ For
example, for wavevectors lying in the \HRS\ one has
\begin{mathletters}
\begin{eqnarray}
\big\langle \qfield{k}\big\rangle_{n+1}^{\rm P} 
&=& 
\big\langle \ofield{k}\big\rangle_{n+1}^{\cal F}\,\,, 
\label{EQ:OPCF}
\\
\big\langle \qfield{k}\,\qfield{k^\prime} 
\big\rangle_{n+1,{\rm c}}^{\rm P}
&=&
\big\langle \ofield{k}\,\ofield{k^\prime} 
\big\rangle_{n+1,{\rm c}}^{\cal F}- 
\frac{V^n}{\mu^2 N}\,\delta_{\hat{k}+\hat{k}^\prime,\hat{0}}\,\,, 
\label{EQ:CORR}
\end{eqnarray}%	
\end{mathletters}%
where $\langle \cdots \rangle_{n+1}^{\cal F}$ denotes an average 
over the field theory~(\ref{EQ:RCMSpartition}), and the subscript 
${\rm c}$ indicates that the correlators are connected.  
Relationships such as those given in Eqs.~(\ref{EQ:OPCF}) and 
(\ref{EQ:CORR}) allow one to relate order-parameter correlators to 
correlators of the field theory. 
%------------------------------------------------------------------
%------------------------------------------------------------------
\subsection{Minimal model for the vulcanization transition}
\label{SEC:LandauWilson}
%------------------------------------------------------------------
The exact field-theoretic representation of RCMSs discussed in the
previous section serves as motivation for a minimal model capable of
describing the universal aspects of the vulcanization transition
inasmuch as it indicates the appropriate order parameter and symmetry
content. In the spirit of the standard Landau approach, one can
determine the {\it form\/} of the minimal model by invoking symmetry
arguments along with three further assumptions: 
%------------
(i)~that fluctuations representing real-space variations in the local 
density of the constituents are free-energetically very costly, and 
should therefore be either suppressed energetically or, equivalently 
(as far as our present aims are concerned), prevented via a kinematic 
constraint; 
%------------
(ii)~that we need only consider order-parameter configurations 
representing physical situations in which the fraction of constituents 
localized is at most small; and (iii)~that the field components 
responsible for the incipient instability of the liquid phase are those 
with long wavelengths.  Provided these assumptions hold, one may: 
%------------
(i)~expand the effective Hamiltonian in powers of the order parameter; 
and 
%------------
(ii)~expand the coefficient functions in powers
of wavevectors. One retains terms only to the order necessary for a
description of both sides of the transition.  (When we go beyond
mean-field theory, below, RG arguments will justify our omission of
all other symmetry-allowed terms on the grounds that they are
irrelevant at the fixed-points of interest.)\thinspace\ This scheme
leads to the following minimal model~\cite{univ,REF:MicroOp}, which 
takes the form of a cubic field theory involving a \HRS\ field 
$\ofield{k}$ that lives on $(n+1)$-fold replicated $d$-dimensional 
space: 
\begin{mathletters}
\begin{eqnarray}
[Z^{n}] 
&\propto&
\int \dmhrs 
\exp( - \action), 
\label{EQ:Partition}
\\
\action
\big(\{\Omega\}\big)
&=&
N\sum_{\hat{k} \in {\HRS}}
\Big(-a\deltat+\frac{b}{2}|\hat{k}|^2\Big)
\big\vert\ofield{k}\big\vert^{2}
-N\inte\,
\sum_{\hat{k}_1,\hat{k}_2,\hat{k}_3\in\HRS}
\oofield{k}{1}\,
\oofield{k}{2}\,
\oofield{k}{3}\,
\delta_{{\hat{k}_1}+{\hat{k}_2}+{\hat{k}_3},{\hat{0}}}\,\,,
\label{EQ:LG_longwave}
\end{eqnarray}
\end{mathletters}
where $\deltat$ is the reduced control-parameter measuring the 
crosslink density.  This model was introduced in Ref.~\cite{univ} as a 
Landau theory of the vulcanization transition, where it was shown to yield 
a rich description of the amorphous solid state, even at the saddle-point 
level, which we briefly summarize in Sec.~\ref{SEC:MeanField} (along with 
the results of various semi-microscopic approaches).  Although the semi-microscopic derivation of $\action$ contains$n$-dependent 
coefficients $a_n$, $b_n$ and $\inte_n$, it is admissible for us to keep 
only the $n\to 0$ limit of these coefficients (i.e.~$a$, $b$ and $\inte$) 
at the outset because $\action$ is already proportional to $n$ for  
pertinent field-configurations .

We wish to emphasize the point that this minimal model {\it does not
contain fields outside the \HRS\/}.  For example, in the cubic 
interaction term in Eq.~(\ref{EQ:LG_longwave}), the wavevectors in the
summations are constrained to lie in the \HRS.  This (linear) constraint 
on the field embodies the notion that inter-particle interactions give a 
\lq\lq mass\rq\rq\ in the \WRS\ (i.e.~produce a free-energy penalty for 
density inhomogeneities) that remains nonzero at the vulcanization 
transition.  From the standpoint of symmetry, this constraint has the 
effect of ensuring that the only symmetry of the theory (associated with 
the mixing of the replicas) is the {\it permutation\/} symmetry 
${\rm S}_{n+1}$.  Without it, the model would have
the larger symmetry, ${\rm O}\left((n+1)d\right)$, of rotations that
mix the (Cartesian components of the) replicas; see the term associated 
with the inter-replica coupling arising from the disorder-averaging of
the replicated crosslinking constraints in Eq.~(\ref{EQ:PureHam}).  In
addition to permutation symmetry, the model has the symmetry of
independent translations and rotations of each replica.  The
restriction to the \HRS\ (or, equivalently, the energetic suppression
of the \WRS) is vital: it entirely changes the content of the theory.
Without it, one would be led to completely erroneous results for both 
the mean-field picture of the amorphous solid state and, as we shall see, 
the critical properties of the vulcanization transition.

For use in Sec.~\ref{SEC:Ginzburg}, when we come to examine the
physical implications of the Ginzburg criterion, we list values of
the coefficients in the action derived for the case of RCMSs (up to
inessential factors of the crosslink density control parameter
$\mu^2$):
\begin{mathletters}
\begin{eqnarray}
\deltat &=& (\mu^2-\mu_{\rm c}^2 )/ \mu_{\rm c}^2, 
\label{EQ:para_t}\\
a &=& 1/2, 
\label{EQ:para_a}\\
b &=& {L\ell}/{6d}, 
\label{EQ:para_b}\\
\inte &=& 1/6.
\label{EQ:para_g}
\end{eqnarray}
\end{mathletters}
Here, $\mu_{\rm c}^2$ is the mean-field critical value of $\mu^2$, 
$L$ is the arclength of each macromolecule, and $\ell$ is the 
persistence length of the macromolecules.
%------------------------------------------------------------------
%------------------------------------------------------------------
\section{Vulcanization transition in 
mean-field theory: Brief summary of results}
\label{SEC:MeanField}
%------------------------------------------------------------------
%------------------------------------------------------------------
\subsection{Mean-field order parameter: 
Liquid and amorphous solid states}
\label{SEC:TLOP}
%------------------------------------------------------------------
Mean-field investigations of RCMSs and related
systems~\cite{epl,cross,univ,REF:endlink,manifolds} have shown that:
%-------------
(i)~There is a continuous phase transition between a liquid and an
amorphous solid state as a function of the density of the crosslinks
(or other random constraints).  This transition is contained within
the \HRS.  Both the liquid and the amorphous solid states have uniform
densities, and therefore the order parameter is zero in the
\WRS\ on both sides of the transition. 
%-------------
(ii)~In the solid state, translational invariance is spontaneously
broken at the microscopic level, inasmuch as a nonzero fraction of the
particles have become localized in space.  However, owing to the
randomness of the localization, this symmetry-breaking is hidden.  [Hence 
the need for a subtle order parameter~(\ref{EQ:opDefinition}).]\thinspace\ 
In the language of replicas, the symmetries of independent translations 
and rotations of the replicas are spontaneously broken, and all that 
remains are the symmetries of common translations and rotations  
(corresponding to the macroscopic homogeneity and isotropy of the 
amorphous solid state).  The permutation symmetry amongst the $n+1$ 
replicas appears to remain intact at the transition.
%-------------
(iii)~The stationarity condition for the order parameter can be 
solved exactly. In the context of the minimal model, in the liquid 
state one finds 
$\mfnot{\Omega}(\hat{k})=0$;   
in the solid state the order parameter takes the form
\begin{mathletters}
\begin{eqnarray}
\mfnot{\Omega}(\hat{k})
&=&
(2a\deltat/3\inte)\,
\delta^{(d)}_{\tilde{\bf k},{\bf 0}}\,\,
\omega\Big(\sqrt{{a\hat{k}^{2}}/{b\deltat}}\,\Big),
\label{EQ:MfResult}
\\
\omega(k)
&\equiv&
\int_{0}^{\infty}d\theta\,\pi(\theta)\,
{\rm e}^{-k^{2}/2\theta}, 
\label{EQ:ord_par_scale}
\end{eqnarray}
\end{mathletters}
where $\tilde{\bf k}\equiv\sum_{\alpha=0}^{n}{\bf k}^{\alpha}$.  The
function $\pi(\theta)$ is a universal function, in the sense that it
does not depend on the model-specific coefficients $a$, $b$ and
$\inte$: it is normalized to unity and satisfies a certain nonlinear
integro-differential equation; see Refs.~\cite{epl,cross,univ}.  From 
the physical perspective, $\omega(k)$ encodes the distribution 
of localization lengths of the localized monomers and the Kronecker 
delta factor $\delta^{(d)}_{\tilde{\bf k},{\bf 0}}$ 
exhibits the macroscopic translational invariance of the random solid 
state.  By passing to the $\hat{k}\to\hat{0}$ limit in 
Eq.~(\ref{EQ:MfResult}) one learns that the fraction of localized 
monomers $q$ (i.e.~the gel fraction) is given by
\begin{equation}
q=\cases{%
         0,			       &liquid state;\cr
	({2a}/{3\inte})\,\deltat^{\beta},&solid  state;}
\label{EQ:MFopRES}
\end{equation}
with the exponent $\beta$ being given by the mean-field value of
unity.  It has recently been demonstrated that the mean-field state
summarized here is locally stable~\cite{REF:stability}.  (We note, in
passing, that no spontaneously replica-symmetry-breaking solutions
of the order-parameter stationary condition have been found, to date.)
%------------------------------------------------------------------
%------------------------------------------------------------------
\subsection{Gaussian correlator: Liquid and critical states}
\label{SEC:TLCO}
%------------------------------------------------------------------
The incipient amorphous solidification, as the vulcanization 
transition is approached from the liquid side, is marked by 
strong order-parameter fluctuations, which are diagnosed via 
the correlator $G(\hat{k})$ defined through  
\begin{equation}
N^{-1}\,\,
\delta_{\hat{k}+\hat{k}^{\prime},\hat{0}}^{(n+1)d}\,\,
G(\hat{k})
\equiv
\big\langle\ofield{k}\,\ofield{k^\prime} 
\big \rangle_{n+1,{\rm c}}^{\cal F}\,\,\,. 
\end{equation}
The unusual factor of $1/N$ is due to our choice of the normalization of 
$\qfield{k}$ in Eq.~(\ref{EQ:DEF_Q}). Section~\ref{SEC:Correlator}, below, 
is dedicated to explaining the physical content of this correlator and
precisely how, via Eq.~(\ref{EQ:CORR}), it is able to detect incipient
random solidification.  The value of the correlator in the mean-field
approximation follows from the quadratic terms in
Eq.(~\ref{EQ:LG_longwave}) and is given by 
\begin{equation}
G(\hat{k})
\approx
G_{0}(\hat{k})
\equiv
{1\over{-2a\deltat+b|\hat{k}|^2}}\,\,, 
\label{EQ:Bare_Cor}
\end{equation}
which below will play the role of the bare propagator.  
Notice that $G_{0}(\hat{k})$ obeys the homogeneity relation 
\begin{equation}
G(\hat{k},\deltat)
\sim 
\vert{\hat k}\vert^{-2+\eta}\,
g\big( \vert\hat{k}\vert\,\vert\deltat\vert^{-\nu} \big), 
\label{EQ:HRphenom}
\end{equation}
in which $g(x)\sim x^{2-\eta}$ for $x\to +0$ 
and approaches a constant value for large $x$.  
Moreover, the exponents take on the mean-field values 
$\eta=0$, 
$\nu=1/2$ 
and
$\gamma=\nu(2-\eta)=1$, 
this last relationship guaranteeing that the susceptibility 
$\lim_{\hat{k}\to\hat{0}}G(\hat{k},\deltat)$ diverges as 
$\vert\deltat\vert^{-\gamma}$. 
%------------------------------------------------------------------
%------------------------------------------------------------------
\section{Order-parameter correlator and susceptibility, 
and their physical significance}
\label{SEC:Correlator}
%------------------------------------------------------------------
Let us now consider the order-parameter correlator 
and the associated susceptibility from the perspective 
of incipient random localization~\cite{REF:ZGprior}.
In the simpler context of, e.g., the ferromagnetic Ising transition
the two-point spin-spin correlator quantifies the idea that the
externally-imposed alignment of a particular spin would induce
appreciable alignment of most spins within roughly one correlation
length of that spin, this distance growing as the transition is
approached from the paramagnetic state.  How are these ideas borne out
in the context of the vulcanization transition?  Imagine approaching
the transition from the liquid side: then the incipient order involves
random localization and so, by analogy with the Ising case, the
appropriate correlator is the one that addresses the question: Suppose
a monomer is localized to within a region of some size by an external
agent: Over what region are other monomers likely to respond by
becoming localized, and how localized will they be? 
We can also consider the order-parameter correlator and the associated
susceptibility from the perspective of the formation of (mobile, thermally 
fluctuating) assemblages of macromolecules, which we refer to as clusters: 
How do they diagnose the development of larger and larger clusters of 
connected macromolecules, as the crosslink density is increased towards 
the vulcanization transition? 

Bearing these remarks in mind, we now examine in detail the physical 
interpretation of the order-parameter correlator 
$\big\langle Q(\hat{k})\,Q(-\hat{k})\big\rangle_{n+1}^{\rm P}$ 
which, as we shall see, captures the physics of incipient localization
and cluster formation.  To see this, consider the construction
\begin{mathletters}
\begin{eqnarray}
C_{\bft}({\bf r}-{\bf r}^{\prime})
\equiv
\bigg[
{V\over{N}}\sum_{j,j^{\prime}=1}^{N}\int_{0}^{1}ds\int_{0}^{1}ds^{\prime}
\big\langle
\delta^{(d)}\big({\bf r}-{\bf c}_{j}(s)\big)\, 
\delta^{(d)}\big({\bf r}^{\prime}-{\bf c}_{j^{\prime}}(s^{\prime})\big)\, 
\big\rangle
\nonumber\\
\qquad\qquad\qquad\qquad
\qquad\qquad\qquad\qquad
\times
\big\langle
\exp-i{\bft}\cdot({\bf c}_{j}(s)-{\bf r})\, 
\exp i{\bft}\cdot({\bf c}_{j^{\prime}(s^{\prime})}-{\bf r}^{\prime})
\big\rangle
\bigg], 
\label{EQ:PhyDef}
\end{eqnarray}%
\end{mathletters}%
which, in addition to depending on the separation 
${\bf r}-{\bf r}^{\prime}$, 
depends on the \lq\lq probe\rq\rq\ wavevector ${\bft}$. The first
expectation value in this construction accounts for the likelihood that
monomers $(j,s)$ and $(j^{\prime}, s^{\prime})$ will respectively be
found around ${\bf r}$ and ${\bf r}^{\prime}$; the second describes
the correlation between the respective fluctuations of monomer $(j,s)$ 
about ${\bf r}$ and monomer $(j^{\prime}, s^{\prime})$ about 
${\bf r}^{\prime}$.

Now, the quantity $C_{\bft}({\bf r}-{\bf r}^{\prime})$ is closely related
to an \HRS\ correlator involving the semi-microscopic order parameter 
$Q(\hat{k})$.  To see this we introduce Fourier representations of the 
two delta functions and invoke translational invariance, thus establishing
that~\cite{REF:MicroRep}
\begin{eqnarray}
C_{\bft}({\bf r}-{\bf r}^{\prime})
&=& 
{N\over V}
\sum\nolimits_{\bf k}
{\rm e}^{\rmi({\bf k}+{\bft})\cdot({\bf r}-{\bf r}^{\prime})}
\Big[
\frac{1}{N^2}
\sum_{j,j^{\prime}=1}^{N}
\int_{0}^{1}ds\,ds^{\prime}\,
\big\langle 
{\rm e}^{-i{\bf k}\cdot
({\bf c}_{j}(s)-
     {\bf c}_{j^{\prime}}(s^{\prime}))}
\big\rangle_\disfac  
\big\langle
{\rm e}^{-i{\bft}\cdot
({\bf c}_{j}(s)-
     {\bf c}_{j^{\prime}}(s^{\prime}))}
\big\rangle_\disfac  
\Big]
\\
&=&
{N\over V}
\sum\nolimits_{\bf k}
{\rm e}^{\rmi{\bf k}\cdot({\bf r}-{\bf r}^{\prime})}
\lim_{n\to 0}\big\langle
Q({\bf 0},{\bf k}-{\bft},{\bft},{\bf 0},\ldots,{\bf 0})^{\ast}\,
Q({\bf 0},{\bf k}-{\bft},{\bft},{\bf 0},\ldots,{\bf 0})
\big\rangle_{n+1}^{\rm P}.
\label{EQ:connect}
\end{eqnarray}

Having seen that $C_{\bft}({\bf r}-{\bf r}^{\prime})$ is closely related
to an \HRS\ correlator involving $Q(\hat{k})$ [which can be computed via 
the $\Omega$ field theory], we now explain in more 
detail how $C_{\bft}({\bf r}-{\bf r}^{\prime})$ detects the spatial 
extent of relative localization.  First, let us dispense with the case of 
${\bft}={\bf 0}$.  In this case $C_{\bft}({\bf r}-{\bf r}^{\prime})$ is 
simply ($V/N$ times) the real-space density-density correlation function 
and, as such, is not of central relevance at the amorphous solidification 
transition.  Next, let us consider the small-${\bft}$ limit of 
$C_{\bft}({\bf r}-{\bf r}^{\prime})$.  This quantity addresses the 
question: If a monomer at ${\bf r}$ is localized \lq\lq by 
hand\rlap,\rq\rq\ what is the likehood that a monomer at 
${\bf r}^{\prime}$ responds by being localized at all, no matter 
how weakly. It is analogous to the correlation function defined in 
percolation theory that addresses the connectedness of 
clusters~\cite{REF:TCL:LH31}.

To substantiate the claim made in the previous paragraph we examine 
the contribution from each pair of monomers to the quantity 
$C_{\bft}({\bf r}-{\bf r}^{\prime})$.  Let us start from the simplest 
situation, in which no crosslinks have been imposed.  We assume that 
${\bft}$ is small (i.e.~$V^{-1/3}\gg\vert\bft\vert^{-1}\gg R_{\rm g}$, 
where $R_{\rm g}$ is the radius of gyration for a single macromolecule) 
and that the macromolecular system has only short-range interactions.  
For each term in the double summation over monomers there are two cases 
to consider, depending on whether or note the pair of monomers are on 
the same macromolecule. For a generic pair of monomers that are on the 
same macromolecule (i.e.~$j=j^\prime$), we expect that 
$\langle\exp i{\bft}\cdot
({\bf c}_{j}(s)-{\bf c}_j(s^{\prime}))\rangle \sim 1$, 
and that (for $\vert{\bf r}-{\bf r}^{\prime}\vert\alt R_{\rm g}$)
$\langle\delta^{(d)}\big({\bf r}-{\bf c}_{j}(s)\big)\, 
        \delta^{(d)}\big({\bf r}^{\prime}-{\bf c}_j(s^{\prime})\big) 
\rangle\sim V^{-1}\,R_{\rm g}^{-d}$.  
Then the total contribution to $C_{\bft}({\bf r}-{\bf r}^{\prime})$ 
coming from pairs of monomers on the same macromolecule is of order 
$(N/V)^{2}R_{\rm g}^{-d}$.  On the other hand, for a generic pair of 
monomers that are on different macromolecules (i.e.~$j\neq j^\prime$), 
we expect that 
$\langle
\exp i{\bft}\cdot
({\bf c}_{j}(s)-{\bf c}_{j^\prime}(s^{\prime}))\rangle
\sim V^{-1}$, 
and that 
$\langle
\delta^{(d)}({\bf r}-{\bf c}_{j}(s))\, 
\delta^{(d)}({\bf r}^{\prime}-{\bf c}_{j^{\prime}}(s^{\prime})) 
\rangle\sim V^{-2}$.  
Therefore the total contribution to $C_{\bft}({\bf r}-{\bf r}^{\prime})$ 
coming from pairs of monomers on different macromolecules is of order 
$(N/V)^{3}V^{-1}$.  Thus, we find that the intrachain (i.e.~$j=j^\prime$) 
contribution to $C_{\bft}({\bf r}-{\bf r}^{\prime})$ dominates over the 
interchain  (i.e.~$j\neq j^\prime$) contribution in the thermodynamic 
limit. 

Moving on to the physically relevant case, in which crosslinks have 
been introduced so as to form clusters of macromolecules, we see that 
what were the intrachain and interchain contributions become intracluster 
and intercluster contributions.  With the appropriate (slight) changes, 
the previous analysis holds, which indicates that the intracluster 
contribution dominates $C_{\bft}({\bf r}-{\bf r}^{\prime})$ in the 
thermodynamic limit.  In other words, in the small-${\bft}$ limit a pair 
of monomers located at ${\bf r}$ and ${\bf r}^{\prime}$ contribute unity 
to $C_{\bft}({\bf r}-{\bf r}^{\prime})$ if they are on the same cluster 
and zero otherwise.  This view allows us to identify the small-${\bft}$ 
limit of $C_{\bft}({\bf r}-{\bf r}^{\prime})$ with the 
pair-connectedness function defined in (the on-lattice version of) 
percolation theory~\cite{REF:TCL:LH31}.

What about $C_{\bft}({\bf r}-{\bf r}^{\prime})$ in the case of general 
${\bft}$?  In this case it addresses the question:  If a monomer near 
${\bf r}$ is localized on the scale $\nbft^{-1}$ (or more strongly), how 
likely is a monomer near ${\bf r}^{\prime}$ to be localized on the same 
scale (or more strongly)?  This additional domain of physical issues 
associated with the strength of localization results from the effects 
of thermal fluctuations, and is present in the vulcanization picture 
but not the percolation one.  

Let us illustrate the significance of $C_{\bft}({\bf r}-{\bf r}^{\prime})$ 
by computing it in the setting of the Gaussian approximation to the 
liquid state in three dimensions.  To do this, we use Eq.~(\ref{EQ:CORR}) 
to express $C_{\bft}({\bf r}-{\bf r}^{\prime})$ in terms of the 
(Gaussian approximation to the) correlator
$\langle\ofield{k}\,\ofield{k^\prime}\rangle_{n+1,{\rm c}}^{\cal F}$, 
which has the Ornstein-Zernicke form given in Eq.~(\ref{EQ:Bare_Cor}).  
Thus, we arrive at the real-space Yukawa form 
\begin{mathletters}
\begin{eqnarray}
\big\vert 
C_{\bft}({\bf r}-{\bf r}^{\prime})
\big\vert 
&\propto&
{\exp\left(
-\vert{\bf r}-{\bf r}^{\prime}\vert/
\zeta_{\rm eff}(\nbft)\right)
\over{\vert{\bf r}-{\bf r}^{\prime}\vert}}, 
\label{EQ:RealSpaceCor} 
\\
{1\over{\zeta_{\rm eff}^{2}(\nbft)}}
&\equiv&
{1\over{\zeta^{2}}}+b\nbft^{2},
\label{EQ:ProbeCorr}
\end{eqnarray}%
\end{mathletters}%
where the correlation length $\zeta$ is defined by 
$\zeta^{-2}\equiv{-2a\deltat}$.  Hence, we see the appearance of a
probe-wavelength--dependent correlation length $\zeta_{\rm eff}(\nbft)$.
The physical interpretation is as follows: in the $\bft\to{\bf 0}$
limit, $C_{\bft}({\bf r}-{\bf r}^{\prime})$ is testing for relative
localization, regardless of the strength of that localization and,
consequently, the range of the correlator diverges at the vulcanization
transition.  This reflects the incipience of an infinite cluster, 
due to which very distant macromolecules can be relatively localized.
By contrast, for generic $\bft$ it is relative localization on a 
scale $\nbft^{-1}$ (or smaller) that is being tested for.  At 
sufficiently large separations, even if a pair of macromolecules are 
relatively localized, this relative localization is so weak that the 
pair does not contribute to $C_{\bft}({\bf r}-{\bf r}^{\prime})$.  
This picture is reflected by the fact that $\zeta_{\rm eff}(\nbft)$ 
remains finite at the transition.

Given that we have identified a correlator that is becoming long-ranged 
at the transition, it is natural to seek an associated divergent 
susceptibility $\Theta_{{\bft}}$.  To do this, we integrate
$C_{\bft}({\bf r}-{\bf r}^{\prime})$ over space and obtain
\begin{equation}
\Theta_{{\bft}}
\equiv
\int
\frac {\rmd^{d}r\,\rmd^{d}r^{\prime}}{V}\,
C_{\bft}({\bf r}-{\bf r}^{\prime})
=
N\lim_{n \to 0}
\left\langle
Q({\bf 0},{\bft},-{\bft},{\bf 0},\ldots,{\bf 0})^{\ast}\,
Q({\bf 0},{\bft},-{\bft},{\bf 0},\ldots,{\bf 0})
\right\rangle_{n+1}^{\rm P}.
\label{EQ:SusceptDef}
\end{equation}
Passing to the ${\bft}\to{\bf 0}$ limit, we have
\begin{equation}
\lim_{{\bft}\to{\bf 0}}\Theta_{{\bft}}
\sim
(-\deltat)^{-\gamma}, 
\label{EQ:suscept}
\end{equation}
where the final asymptotic equality is obtained from a 
computation of the (field-theoretic) correlator
$\big\langle\ofield{k}\,\ofield{k^\prime}
 \big\rangle_{n+1,{\rm c}}^{\cal F}$ 
[see Eq.~(\ref{EQ:CORR})]. 
This quantity is measure of the spatial extent over which pairs of 
monomers are relatively localized, no matter how weakly, and thus 
diverges at the vulcanization transition.  At the Gaussian level of 
approximation, Eq.~(\ref{EQ:Bare_Cor}), this susceptibility diverges 
with the classical exponent $\gamma=1$.  By contrast, for generic 
$\bft$ the susceptibility $\Theta_{{\bft}}$ remains finite at the 
transition, even though an infinite cluster is emerging, due to the 
suppression of contributions to $\Theta_{{\bft}}$ from pairs of 
monomers whose relative localization is sufficiently weak   
(i.e.~those that lead to the divergence in the small-${\bft}$ limit).
%------------------------------------------------------------------
%------------------------------------------------------------------
\section{Vulcanization transition beyond mean-field theory}
\label{SEC:BeyondTree}
%------------------------------------------------------------------
%------------------------------------------------------------------
\subsection{Ginzburg criterion for the vulcanization transition}
\label{SEC:Ginzburg}
%------------------------------------------------------------------
To begin the process of analyzing the vulcanization transition beyond
the mean-field (i.e.~tree) level, we estimate the width
$\delta\deltat$ of reduced constraint-densities $\deltat$ within which
the effects of order-parameter fluctuations about the saddle-point
value cannot be treated as weak, i.e., we construct the Ginzburg
criterion.  To do this, we follow the conventional strategy (see,
e.g., Ref.~\cite{Amit}) of computing a loop expansion for the 2-point
vertex function to one-loop order and examining its low-wavevector
limit (i.e.~the inverse susceptibility).  Note that in the present
setting the loop expansion amounts to an expansion in the inverse
monomer density.

%--------------------------
\begin{figure}[hbt]
\vskip0.50cm
\epsfxsize=3.5in
  \centerline{\epsfbox{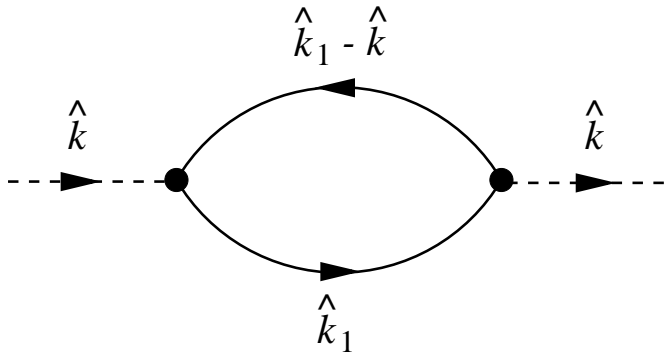}} 
% \centerline{\epsfbox{fig1.ps}} 
\vskip0.50cm
\caption{One-loop correction to the 2-point vertex function.  
Full lines indicate bare \HRS\ correlators; 
dashed lines indicate amputated external bare \HRS\ correlators.
\label{FIG:ginzburg}}
\end{figure}
%--------------------------
Our starting point is the minimal model, Eq.~(\ref{EQ:LG_longwave}), 
for which the bare
correlator is given by Eq.~(\ref{EQ:Bare_Cor}). Then the one-loop
correction to the 2-point vertex function comes from the diagram shown
in Fig.~\ref{FIG:ginzburg}, which is calculated in
App.~\ref{SEC:GC_diagrams}. By choosing $\hat k \in 3^+RS$ (i.e.~in the 
\HRS\ but not in the two-replica sector)~\cite{REF:sector_disc} we obtain
for the inverse susceptibility $\chi^{-1}$ the result
\begin{equation}
(N\chi)^{-1} = 
-2a\deltat+
18\inte^2 \frac {V}{N} \int \frac {d^d p}  
	{ (-2a\deltat +b p^2)^2}\,\,,  
\label{EQ:preshift}
\end{equation}
in which a large wavevector cut-off at $\vert\hat{k}\vert=\Lambda$ is
implied.  The (one-loop) shifted critical point $\deltat_{\rm c}$ 
marks the vanishing of $\chi^{-1}$, i.e., solves
\begin{equation}
0 = -2a\deltat_{\rm c} + 18\inte^2  \frac {V}{N} 
	\int \frac {d^d p} {(-2a\deltat_{\rm c} +b p^2)^2}\,\,. 
\label{EQ:shift}
\end{equation}
Now, in mean-field theory the transition occurs at $\deltat=0$, with
positive (resp.~negative) values corresponding to the amorphous solid
(resp.~liquid) states.  From Eq.~(\ref{EQ:shift}) we see that that
inclusion of fluctuations enlarges the region of crosslink densities
in which liquid state is stable, as one would expect on general
physical grounds.  However, it is worth noting, in passing, that without
the exclusion of the one-replica sector the converse would occur
(i.e.~fluctuations would enlarge the region of stability of the
amorphous solid state). By subtracting Eq.~(\ref{EQ:shift}) from
Eq.~(\ref{EQ:preshift}) in the standard way, replacing $\deltat_{\rm
c}$ by its mean-field value (of zero) in the loop correction, and
rescaling the integration variable $p^2$ according to $ b p^2 =
-2a\deltat k^2$, we arrive at
\begin{equation}
(N\chi)^{-1}
=-2a(\deltat-\deltat_c) 
	\left( 1- 18 \inte^2 ({V}/{N})  b^{-d/2} 
	(-2a\deltat)^{(d-6)/2} J_d \right), 
\label{EQ:Ginz_Crit}
\end{equation}
where $J_d$ is a dimensionless number dependent on $d$ (and weakly on
$\Lambda$, at least in regime of interest, i.e., $d$ below 6).
Equation~(\ref{EQ:Ginz_Crit}) shows that for $d<6$ a fluctuation
dominated-regime is inevitable for sufficient small $\deltat$, and
hence that the upper critical dimension for the vulcanization
transition is six, in agreement with na{\"\i}ve power-counting
arguments applied to the $n\to 0$ limit of the cubic field theory,
Eq.~(\ref{EQ:LG_longwave}).  The Ginzburg criterion amounts to
determining the departure of $\deltat$ from its critical value such
that in Eq.~(\ref{EQ:Ginz_Crit}) the one-loop correction is comparable
in magnitude to the mean-field level result.

To determine the physical content of the Ginzburg criterion, we invoke
the values of the coefficients of the minimal model appropriate for
the semi-microscopic model of RCMSs,
Eqs.~(\ref{EQ:para_t})-(\ref{EQ:para_g}), and we exchange the
macromolecule density $N/V$ for the volume fraction
$\volfrac\equiv(N/V)(L/\ell)\ell^{d}$.  Thus we arrive at the
following form of the Ginzburg criterion: for $d<6$, fluctuations
cannot be neglected for values of $\deltat$ satisfying
\begin{equation}
\vert\deltat-\deltat_{\rm c}\vert
\lesssim 
\left({L/{\ell}}\right)^{-\frac{d-2}{6-d}}
\left({\volfrac/{\inte^{2}}}\right)^{-\frac{2}{6-d}}, 
\end{equation}
from which we see that the fluctuation-dominated regime is narrower 
for longer macromolecules and higher densities (for $2<d<6$).  
Such dependence on the degree of polymerization $L/\ell$ is precisely 
that argued for long ago by de~Gennes on the basis of a 
percolation-theory picture~\cite{REF:DeGennes}. 

Besides the fields and vertices featuring in the minimal model, there
are other symmetry-allowed fields and vertices that are generated by the 
semi-microscopic theory of RCMSs. Examples are provided by the \WRS\ field, 
which describes density fluctuations, along with vertices of cubic, quartic 
or higher-order that couple the \WRS\ field to the \HRS\ field. In
App.~\ref{SEC:Subleading} we investigate the effect of these fields
and vertices, which are omitted from the minimal model, and show:
(i)~that the inclusion of their effects (at the one-loop level) does
not change the Ginzburg criterion derived in the present section; and
(ii)~that the \HRS\ critical fluctuations do not provide any singular
contributions to the \WRS\ density-density correlation function (at
least to one-loop order).
%------------------------------------------------------------------
%------------------------------------------------------------------
\subsection{Renormalization-group procedure and its subtleties}
\label{SEC:RG_philosophy}
\begin{figure}[hbt]
 \vskip0.50cm
 \epsfxsize=3.5in
  \centerline{\epsfbox{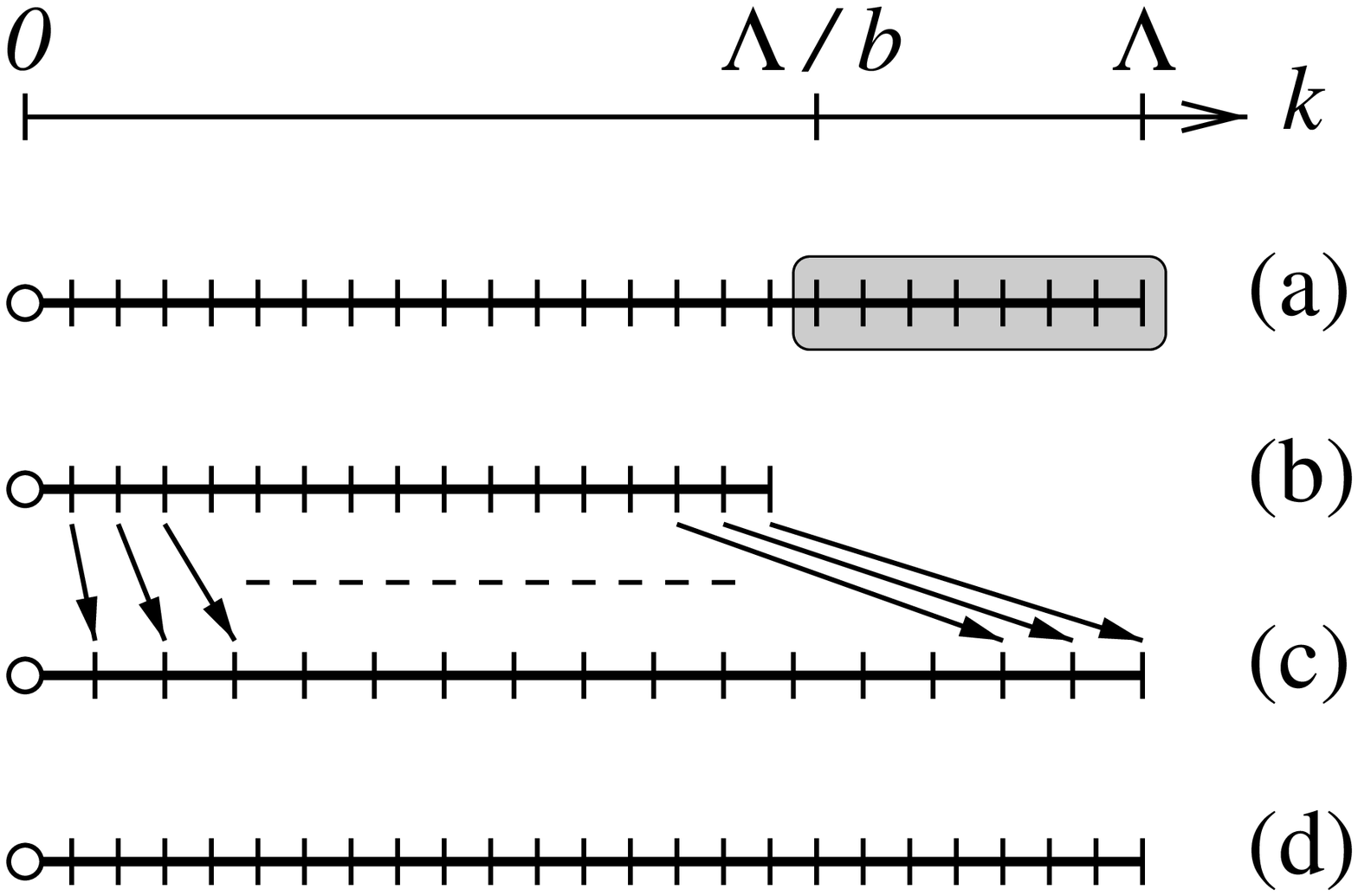}} 
% \centerline{\epsfbox{fig2.ps}} 
\vskip0.75cm
\caption{Schematic one-dimensional depiction of the basic steps 
of the RG procedure (the field variables are defined only at the
hash marks denoting the quantized wavevectors): from (a) to
(b)~integrate out the fields at the quantized wavevectors $k$ in the
\lq\lq momentum shell\rq\rq  (shaded); from (b) to (c)~rescale lengths
to restore the wavevector cut-off, and rescale the field to restore
the gradient term; from (c) to (d)~restore the density of the degrees
of freedom.  (In practice, we employ a momentum shell of infinitesimal
width.)}
\label{FIG:coarse}
\end{figure}

%------------------------------------------------------------------
We now describe the RG procedure that we are using, a schematic 
depiction of which is given in Fig.~\ref{FIG:coarse}. 
The main thrust of our approach is the standard 
\lq\lq momentum-shell\rq\rq\ RG, via which we aim to determine 
how the parameters of the theory, $\deltat$ and $\inte$, flow under
the two RG steps of coarse graining and rescaling.  However, in the
present context there are some significant subtleties owing to the need
to constrain the fields to lie in the \HRS.

In the coarse-graining step, we integrate out the rapidly-varying
components of $\ofield{k}$ (i.e.~those corresponding to wavevectors
satisfying $\Lambda/b<\vert\hat{k}\vert<\Lambda$).  Here, the
constraint that only the \HRS\ field is a critical field demands that
one treat the \HRS\ and the \WRS\ distinctly.  We handle this by
working with a large but finite (replicated) system contained in a
hyper-cubic box of volume $V^{n+1}$ on which periodic boundary
conditions are applied.  As a consequence, the wavevectors are \lq\lq
quantized\rlap,\rq\rq\ and therefore we can directly make the
appropriate subtractions associated with the removal of the zero- and
one-replica sectors.  Having made the necessary subtractions, we
compute the various Feynman diagrams (for the construction of the
Ginzburg criterion and the the coarse-graining step of the RG) by
passing to the continuous wavevector limit (so that wavevector
summations become integrations).

The replica technique has the following curious feature. 
In the infinite-volume limit the different sectors are spaces of 
different dimensionalities, and thus the contributions from the 
lower replica sectors appear to be sets of measure zero relative 
to the contributions from the \HRS.  However, in the replica limit, 
the contributions from different sectors are comparable and, hence, 
the lower sectors cannot be neglected. 

%--------------------------
%--------------------------
The coarse-graining step is followed by the rescaling step, in which
the aim is to return the theory to its original form.  The field- and
length-rescaling aspects of this step (to recover the original
wavevector cut-off and form of the gradient term) are standard, but
there is a subtlety associated with the fact that the original theory
is defined on a finite volume (in order that the wavevectors be
quantized and the various replica sectors thereby be readily
identifiable).  This subtlety is that upon coarse-graining and rescaling
one arrives at a theory that is {\it almost\/} of the original form,
but is defined on a coarser lattice of quantized wavevectors
associated with the reduced (real-space) volume.  If we wish to return
the theory to its truly original form, we are required to increase the
density of the coarsened wavevector lattice.  To accomplish this, we
choose to make use of the extension to $(n+1)d$ dimensions of the
following one-dimensional relation, exact in the thermodynamic
(i.e.~large real-space size $B$) limit:
\begin{eqnarray}
\sum_{k\in\{2\pi nb/B\}}f(k)
\approx
{b}^{-1}\sum_{k\in\{2\pi n/B\}}f(k). 
\label{EQ:thin}
\end{eqnarray}
One way to understand this is to regard the two sides of
Eq.~(\ref{EQ:thin}) as providing different discrete approximations to
the same continuous-wavevector (i.e.~infinite-volume) limit.  Thus, we
expect the difference between them to be unimportant in the
thermodynamic limit.  Another way is to regard the right hand side of
Eq.~(\ref{EQ:thin}) as pertaining to a system with a larger number of
degrees of freedom than the left hand side, but that the factor
$b^{-1}$ appropriately diminishes the weight of each degree of
freedom.  It would be equally satisfactory if we chose, in our 
RG scheme, {\it not\/} to restore the wavevector lattice spacing, 
which would amount to our using the left-hand-side of Eq.~(\ref{EQ:thin}).
%------------------------------------------------------------------
%------------------------------------------------------------------
\subsection{Expansion around six dimensions}
\label{SEC:EpsilonExpansion}
%------------------------------------------------------------------
In the previous two subsections we have established that the upper
critical dimension for the vulcanization transition is six, and we
have described an RG procedure capable of elucidating certain
universal features of the transition.  We now examine the RG flow
equations near the upper critical dimension that emerge from this
procedure, and discuss the resulting fixed-point structure and
universal critical exponents.  To streamline the presentation we have
relegated the technical details of the derivation of the flow
equations to App.~\ref{SEC:RG_floweq}.
%------------------------------------------------------------------
%------------------------------------------------------------------
\subsubsection{Flow equations}
\label{SEC:FlowEpsilonExpansion}
%------------------------------------------------------------------
As with the mean-field theory and the Ginzburg criterion, our starting
point is the replicated cubic field theory,
Eq.~(\ref{EQ:LG_longwave}).  By suitably redefining the scales of
$\ofield{k}$ and $\hat{k}$ we can absorb the coefficients $a$ and $b$,
hence arriving at the Landau-Wilson effective Hamiltonian
\begin{equation}
\action\big(\{ \Omega \} \big) = 
N \sum_{\hat{k} \in {\HRS}}
\Big(-\deltat+\frac{1}{2}|\hat{k}|^2\Big)
\big\vert\ofield{k}\big\vert^{2}
-N\inte\,
\sum_{\hat{k}_1,\hat{k}_2,\hat{k}_3\in {\HRS}} 
\oofield{k}{1}\,
\oofield{k}{2}\,
\oofield{k}{3}\,
\delta_{{\hat{k}_1}+{\hat{k}_2}+{\hat{k}_3}, {\hat{0}}}\,, 
\label{EQ:RG_ft}
\end{equation} 
in which all wavevector summations are cut off beyond 
replicated wavevectors of large magnitude $\Lambda$, 
from which we can read off the bare correlator 
\begin{equation}
G_0(\hat k)=\frac{1}{-2\deltat+\vert\hat{k}\vert^2}\,\,. 
\end{equation} 

%--------------------------
\begin{figure}[hbt]
 \epsfxsize=4.0truein 
  \centerline{\epsfbox{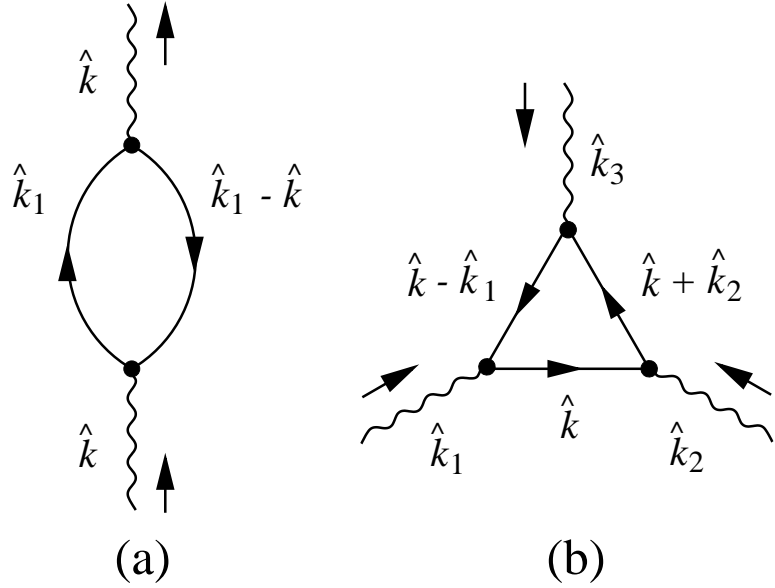}} 
% \centerline{\epsfbox{fig3.ps}} 
 \vskip0.50cm
 \caption{Contributing one-loop diagrams.  Full lines indicate 
bare \HRS\ correlators for short-wavelength fields (i.e.~fields 
lying in the momentum shell); wavy lines indicate long-wavelength 
fields.
 \label{FIG:renorm}}
\end{figure}%
%--------------------------
We shall be working to one-loop order and, correspondingly, the 
diagrams that contribute to the renormalization of the parameters 
of the Landau-Wilson effective Hamiltonian are those depicted in 
Figs.~\ref{FIG:renorm}~(a) and (b). The resulting flow equations are
%------------------
\begin{mathletters}
\begin{eqnarray}
d\deltat/d\ln b
&=& 
	2\,\deltat -C_0\, \inte^2- C_0^{\prime}\,\deltat\, \inte^2 - 
	C_1\, \deltat \,\inte^2 + 
	{\cal O}(\deltat^2\inte^2,\deltat\wpeps\inte^2,
		\wpeps\inte^2, \inte^4),
\label{EQ:ctmflowdel}
\\
d\inte/d\ln b
&=& 
	\inte\big( 
		\wpeps/2 - C_3\inte^2 -\frac{3}{2}C_1 \inte^2 + 
		{\cal O}(\deltat \inte^2, \wpeps \inte^2, \inte^4)
	\big), 
\label{EQ:ctmflowint}
\\
dz/d\ln b
&=& 
	{1\over{2}}(d+2-C_{1}\inte^{2})
		+{\cal O}(\deltat \inte^2, \wpeps \inte^2, \inte^4),
\label{EQ:ctmflowzee}
\end{eqnarray}
\end{mathletters}
%------------------
where $\wpeps\equiv 6-d$, $b$ is the length-rescaling factor, $z$ is
the field-rescaling factor, and the (constant) coefficients in the flow
equations are given by
\begin{equation}
(C_0, C_0^{\prime}, C_1, C_3)= 
	\frac{V}{N} \frac {S_6}{(2\pi)^6}\, 
	( 9\Lambda^2, 36, -6, 72), 
\label{EQ:FlowCoe}
\end{equation} 
in which $S_6$ is the surface area of a $6$-dimensional sphere 
of unit radius.
%------------------------------------------------------------------
%------------------------------------------------------------------
\subsubsection{Fixed-point analysis and its consequences}
\label{SEC:FixedEpsilonExpansion}
%------------------------------------------------------------------
We proceed in the standard way by first finding the fixed points
$(\deltat_*, \inte_*)$ of the flow equations, at which $d(\deltat,
\inte)/d\ln b = (0,0)$.  We linearize the flow equations about each of
the resulting fixed points,
\begin{equation}
{d\over{d\ln b}}
\pmatrix{
\deltat-\deltat_{*}\cr
\inte-\inte_{*}}
\approx
\pmatrix{
2-\big(C_0^{\prime}+C_1\big)\inte_{*}^{2}& 
-2C_0\,\inte_{*}
\cr
0&
\frac{1}{2}\wpeps-3\big(C_3+\frac{3}{2}C_1\big)\inte_{*}^{2}
\cr}
\pmatrix{
\deltat-\deltat_*\cr
\inte-\inte_*}, 
\label{EQ:LinFlowEq}
\end{equation}
where we have dropped higher-order corrections. We then establish
the RG eigenvalues at each fixed point by finding the eigenvalues of 
the linearized RG transformation matrices. Finally, we solve
Eq.~(\ref{EQ:LinFlowEq}) to obtain the flow near each fixed point.

For $\wpeps$ both negative and positive (i.e.~for $d$ both above 
and below six) we find a {\it Gaussian\/} fixed point (GFP): 
$(\deltat_*, \inte_*)=(0, 0)$.  Solving Eq.~(\ref{EQ:LinFlowEq}) 
about this fixed point gives the flow 
\begin{equation}
\pmatrix{
\deltat(b)\cr
\inte(b)}
\approx
\pmatrix{
\deltat(1)\,b^{y_{1}}\cr
\inte(1)  \,b^{y_{2}}}, 
\label{EQ:LinFlowGFP}
\end{equation}
with the RG eigenvalues $y_{1}$ and $y_{2}$ respectively given by 
$y_{\deltat}=2$ and $y_{\inte}=\wpeps/2$.

As one can see from Eq.~(\ref{EQ:LinFlowGFP}), above six dimensions 
the GFP is unstable in the $\deltat$ direction and stable in the
$\inte$ direction.  However, below six dimensions the GFP also
becomes unstable in the $\inte$ direction, and a new fixed point---the
Wilson-Fisher fixed point (WFFP)---emerges, located at
$(\deltat_*,\inte_*^2)=
\left((\Lambda^2/28),(1/126)((2\pi)^6/ S_6)
(V/N)^{-1}\right)\wpeps$. 
(Let us mention, in passing, that if we had not correctly implemented the 
constraint that wavevector summations exclude contributions for the 
\WRS\ then the structure of the flow equations would have been utterly
different; e.g., the WFFP would have occurred at a complex value of 
$\inte$.)\thinspace\ By solving Eq.~(\ref{EQ:LinFlowEq}) for the WFFP 
we find the flow 
\begin{equation}
\pmatrix{
\deltat(b)-\deltat_{*}\cr
\inte(b)-\inte_{*}}
\approx
\pmatrix{
 \big(\deltat(1)-\deltat_{*}\big)-
 A\big(\inte(1)-\inte_{*}\big)
	\cr
0}b^{y_{1}}+
\pmatrix{
 A\big(\inte(1)-\inte_{*}\big)
	\cr
\inte(1)-\inte_{*}}b^{y_{2}}, 
\label{EQ:LinFlowWFFP}
\end{equation}
where 
$A\equiv
 \big(3/\sqrt{14}\big)
 \big((V/N)(S_6/(2\pi)^6)\big)^{1/2}
 \big(\Lambda^{2}\wpeps^{1/2}\big)$ 
and  the RG eigenvalues are given by $y_{1}=2-(5\wpeps/21)$ and 
$y_{2}=-\wpeps$. 

We now proceed to obtain the critical exponents for physical quantities 
from the RG eigenvalues at each fixed point.  The homogeneity relation 
for the correlator $G({\hat k})$, following from a standard RG
analysis~\cite{REF:CardyEsp}, reads
\begin{equation}
G({\hat k},\deltat)
=z^{2}\,b^{-d}\,
G(b{\hat k},b^{y_{1}}\deltat).
\label{EQ:GScale}
\end{equation}
We eliminate $b$ by choosing $b\vert{\hat k}\vert=1$; then comparison 
with Eq.~(\ref{EQ:HRphenom}) leads to  $\nu=1/y_{1}$ and
$\eta=C_{1}\inte_{\ast}^{2}$. Thus, for the GFP we have
\begin{equation}
\nu^{-1}=2,\qquad
\eta=0,
\end{equation}
and for the WFFP we have, to first order in $\wpeps$, 
\begin{equation}
\nu^{-1}=2-(5\wpeps /{21}),\qquad
\eta=     -{\wpeps/{21}}.
\end{equation}
Both above and below six dimensions, the critical exponents $\nu$ and 
$\eta$ (and $\beta$, to be discussed below) are identical to those 
governing analogous quantities in percolation theory (at least to first 
order in $\wpeps$), as computed via the Potts field theory~\cite{Harris}.  
We discuss the significance of this result and the relationship between 
the present approach and percolation/gelation-based approaches in 
Sec.~\ref{SEC:Potts}.

We have focused on the cubic interaction in the vulcanization field 
theory.  There are, of course, additional symmetry-allowed 
interactions, such as the quartic interaction.  Near to six dimensions, 
however, the fact that such interactions are irrelevant at the GFP 
can be shown by na{\"\i}ve power-counting arguments, which hold in 
the replica limit (and remain uncompromised at the WFFP, owing to its 
proximity to the GFP). 
%------------------------------------------------------------------
%------------------------------------------------------------------
\subsection{Scaling for gel fraction and wavevector-dependent order parameter}
\label{SEC:Scaling}
%------------------------------------------------------------------
In order to relate properties of the amorphous solid state to those
computed in the liquid and critical states, we now follow
the standard scaling analysis.  To do this, we add to the minimal
model, Eq.~(\ref{EQ:RG_ft}), a source field that couples linearly to the
order parameter: $-N\sum_{\hat{k}\in{\HRS}}\ofield{k}\,U(-\hat k)$. We
assume that $U$ contains only long wavelength components, so that it
does not couple to any field featuring in any momentum-shell
integrations.  Then the renormalization of $U(-{\hat k})$ comes only
from the rescalings of $\hat k$ and $\Omega({\hat k})$, and thus we
have
\begin{equation}
U^{\prime}({\hat k}^{\prime})= 
z\,b^{-d}\,U({\hat k}). 
\end{equation}
To obtain the exponent $\beta$, which describes the scaling of the gel 
fraction $q$, the conventional method prescribes the application of a 
{\it uniform\/} source field.  In the present theory, the (zero replica 
sector) field variable $\Omega({\hat 0})$, which would couple to such
a uniform source, is excluded, and instead we choose 
$U({\hat k})=h\,\delta_{{\hat k}+{\hat k}_0,{\hat 0}}$, 
where ${\hat k}_0$
lives in the \HRS\ but is otherwise arbitrarily small.  (This
prescription is consistent with the notion that the gel fraction
follows from the long-wavelength limit of the order parameter, the
limit being taken via wavevectors in the \HRS.)\thinspace\ Hence we
arrive the recursion relation for $h$:
\begin{equation}
h^{\prime} =z\,h=b^{y_h};\qquad y_h=(d+2-\eta)/2. 
\end{equation} 
As we are already in possession of $\eta$ at the GFP and the WFFP, we
thus arrive at the scaling dimension $y_h$ of the source field $h$.

Having obtained $y_h$, we now use it, together with $y_\deltat$, 
$y_\inte$ and the singular part of the free energy density $f$, to 
determine $\beta$, in the following way.  According to homogeneity, 
$f$ has the form
\begin{equation}
f(\deltat, \inte, h) = 
b^{-d} f( \deltat b^{y_\deltat},\inte b^{y_\inte}, hb^{y_h}). 
\end{equation}
By taking the derivative with respect to $h$ so as to form the
order-parameter equation of state, choosing $h=0$, and passing to the
small ${\hat k}_{0}$ limit, one finds the following scaling behavior
of the gel fraction:
\begin{equation}
q(\deltat,\inte,0)\sim
\lim\limits_{\hat{k}_0\rightarrow\hat{0}}
\partial f/\partial h\big\vert_{h=0}\sim
b^{-d+y_h}\,M(\deltat b^{y_\deltat}, \inte b^{y_\inte}, 0)
	= \deltat^{{(d-y_h)}/{y_\deltat}} 
	  M(1, \inte\deltat^{-y_\inte/y_\deltat}, 0).
\end{equation}

Let us first consider the regime $d>6$, for which the appropriate fixed 
point is the GFP and, therefore one expects the exponents to take on their 
classical values.  Now, as one can see from the mean-field value for the 
order parameter $\mfnot{\Omega}$ (and thus the gel fraction $q$), 
Eq.~(\ref{EQ:MfResult}), both of which are proportional to $\inte^{-1}$, the 
cubic interaction is dangerously irrelevant at the GFP, and thus one has 
\begin{equation}
M(1, \inte, 0) \sim \frac{1}{\inte}, 
\quad{\rm for}\quad 
\inte \rightarrow +0.
\end{equation}
Hence, near the GFP one has 
\begin{mathletters}
\begin{eqnarray}
&&\qquad\qquad\qquad
q(\deltat,\inte,0)
\sim
\deltat^{\beta},
\quad{\rm for}\quad 
\inte\rightarrow +0,
\\
&&\qquad\qquad\qquad
\beta
=\frac{d-y_h}{y_\deltat}
+\frac{y_\inte}{y_\deltat}
=\frac {d-\frac{d+2}{2}+\frac{6-d}{2}}{2} =1,
\end{eqnarray}%
\end{mathletters}%
which is precisely the mean-field value of the exponent $\beta$ given
in Sec.~\ref{SEC:TLOP}.

Now let us turn to the regime $d<6$, for which the exponents are 
nonclassical.  The appropriate fixed point is now the WFFP, at 
which the cubic interaction is irrelevant but not dangerously so.  
Thus, in this regime one has the standard scaling relation
\begin{equation}
\beta=\frac{d-y_h}{y_\deltat}=1-(\wpeps/7),
\end{equation}
where the second equality holds only to order $\wpeps$.

In fact, under the (not unreasonable) assumption that there is only 
one characteristic length-scale in the ordered state (i.e.~that the 
fluctuation correlation length does not provide a length-scale 
independent from the localization length-scale), we can go beyond 
the establishing of the scaling of the gel fraction (i.e.~the 
long-wavelength limit of the order parameter) and propose a more general 
scaling hypothesis, which incorporates the scaling of the (singular 
part of the) wavevector-dependent order parameter~\cite{REF:condense}.  
This takes the form of the scaling hypothesis: 
\begin{equation}
\big\langle\ofield{k}\big\rangle 
\propto 
\deltat^{\beta}\,w\big({\hat k}^2 \deltat^{2\nu}\big).
\end{equation}
The quantity $\deltat^{\nu}$, which plays the role of the 
fluctuation correlation length in the liquid state, is here 
seen to play the role of the characteristic scale for the 
localization lengths in the ordered state.  Presumably, it 
also governs the scale over which (amplitude-type)  
fluctuations are correlated in the solid state.  Let us note 
that the mean-field result for the order parameter not only 
obeys this scaling relation (with $\beta=2\nu=1$) but also 
provides an explicit form for the function $w$.
%------------------------------------------------------------------
%------------------------------------------------------------------
\section{Concluding remarks: Connections with 
other approaches and the role of thermal fluctuations}
\label{SEC:Potts}
%------------------------------------------------------------------
%------------------------------------------------------------------
Having constructed an RG theory for the liquid and critical states 
of vulcanized matter, we now examine the results of this RG theory  
and discuss the relationship between these results and the results 
of other approaches to the vulcanization transition.
As we have seen in Secs.~\ref{SEC:EpsilonExpansion}, via an expansion
around six spatial dimensions our minimal model for the vulcanization
transition yields values for certain critical exponents that
characterize the behavior of the system near to and at the transition.
These exponents turn out to be numerically equal to those
characterizing physically analogous quantities in percolation theory,
at least to first order in the departure $\wpeps$ from six dimensions.
We have not proven that the equality between exponents holds beyond
first order in $\wpeps$, although there are hints in the structure of
the theory suggesting that it does.

This equality between exponents seems reasonable in view of the intimate 
relationship between percolation theory and the {\it connectivity\/}
of the system of crosslinked macromolecules, this connectivity
pertaining to the {\it statistics\/} of systems formed according to
the Deam-Edwards distribution of quenched randomness (and hence to the
statistical mechanics of the uncrosslinked macromolecular
liquid)~\cite{REF:ButQRV}.  Indeed, a connection between the
percolation and vulcanization transitions already shows up at the
level of mean-field theory: the dependence of the gel fraction $q$ on
the crosslink-density control parameter $\mu^{2}$ obtained via the
semi-microscopic approach (in the case of RCMSs), viz., that $q$ obeys
\begin{equation}
1-q=\exp\left(-\mu^{2}q\right),
\end{equation}
is identical to the mean-field--percolation dependence of the fraction
of sites participating in the infinite cluster, obtained by Erd{\H o}s
and R\'enyi in their work on random graphs~\cite{REF:ER}, this
identity holding not just near the transition, where the dependence of
$q$ on $\mu^{2}-\mu_{\rm c}^{2}$ is linear, but for all crosslink
densities.  Moreover, the mean-field result emerging from the minimal
model of the vulcanization transition 
yields this linear dependence (but cannot, of course, be applied
beyond the transition regime).  The relevance of percolation theory to
the vulcanization transition also manifests itself beyond the
mean-field level in the physical {\it meaning\/} of the order-parameter
correlator, as we have discussed in Sec.~\ref{SEC:Correlator}.  This
connection has long been realized, and supports the use of percolative
approaches as models of certain aspects of the vulcanization transition~\cite{REF:FloryBook,REF:Stauffer,REF:PGDGbook,REF:StCoAd,REF:LubIsaac}.

These percolative approaches include direct applications of percolation
theory~\cite{REF:FloryBook,REF:Stauffer,REF:PGDGbook,REF:StCoAd}, 
mentioned in the preceding paragraph, as well as the approach given  
by Lubensky and Isaacson~\cite{REF:LubIsaac}.  The latter approach 
extends the connection between the statistics of linear macromolecules 
and the zero-component limit of a spin 
system~\cite{REF:NGZtrick,REF:JDCfields}.  
In this way, a correspondence is established between the statistics 
of branched, polydisperse, macromolecules and a multi-component field 
theory.  This field theory reduces to the one-state limit of the Potts
model under circumstances appropriate for the transition to a physical 
gel (i.e.~a state in which one is certain to find a reversibly-bonded,  
infinite, branched macromolecule)~\cite{REF:roleQR}. 

An essential ingredient of the approaches discussed in the previous
paragraph is the Potts model in its one-state limit---a representation
of percolation~\cite{REF:FortKast,REF:CardyPerc}.  It is
therefore worth considering similarities and differences between the
minimal field theory of the vulcanization transition focused on in the
present Paper, Eq.~(\ref{EQ:LG_longwave}), and the minimal field theory
for the Potts model. The minimal field theory for the Potts model is
the $n\to 0$ limit of the cubic $(n+1)$-component field theory, the
Landau-Wilson Hamiltonian for which is
\begin{equation}
\int_{V}\rmd^{d}x\,
        \Big(
\,\sum_{\alpha=1}^{n}
\left(
 {1\over{2}}r\,\psi_{\alpha}^{2}
+{1\over{2}}\vert\bnabla\psi_{\alpha}\vert^{2}
\right)
-w^{(3)}\sum_{\alpha,\beta,\gamma=1}^{n}
 \lambda^{(3)}_{\alpha\beta\gamma}\,
 \psi_{\alpha}\,
 \psi_{\beta}\,
 \psi_{\gamma}
        \Big),
\label{EQ:Potts}
\end{equation}
where $r$ controls the bond-occupation probability (and hence the
percolation transition), $w^{(3)}$ is the nonlinear coupling strength, and
$\lambda^{(3)}_{\alpha\beta\gamma}$ is the \lq\lq Potts tensor\rq\rq\
(which controls the internal symmetry of the theory; for a discussion
of this theory see, e.g., Sec.~2.7 of Ref.~\cite{REF:TCL:LH31}).  

How does this Potts field theory compare the vulcanization field theory 
that we have been analyzing in the present Paper?  The Potts field 
theory has, a cubic interaction, as does the vulcanization field theory, 
and therefore its upper critical dimension is also six.  If we examine 
the RG analysis of the Potts field theory (in an expansion around six
dimensions)~\cite{REF:Amit_Paper} we see that, at the one-loop level, 
diagrams identical in form (i.e.~those shown in Fig.~\ref{FIG:renorm}) 
enter the renormalization of the various vertices.  
%---
Moreover, in the $n\to 0$ limit the RG flow equations for the two theories 
turn out to be identical.  This striking result is connected to the 
following observations: 
%---
\hfil\break\noindent
(i)~In Potts case, aside from the $d$-dimensional integrals corresponding 
to the diagrams, the coefficients in the flow equations are determined by 
the contractions of Potts-tensor indices associated with each cubic vertex, 
these contractions being the origin of the $n$-dependence of the 
coefficients in the flow equations. 
%---
\hfil\break\noindent
(ii)~In the vulcanization case, the diagrams intrinsically correspond to 
$(n+1)d$-dimensional integrals but, due to the constraints on the
summations over wavevectors, these diagrams produce $(n+1)d$-dimensional
integrals (which smoothly reduces to $d$-dimensional integrals in the
$n\to 0$ limit), together with $d$-dimensional integrals 
[see Eqs.~(\ref{EQ:RG_Cons_Sum_3+},\ref{EQ:FtwoInt})]. 
%---
\hfil\break\noindent
(iii)~Despite the explicit differences in the forms of the two theories, 
it turns out that, in the $n\to 0$ limit, the integrals and the 
combinatorics conspire to produce precisely the same flow equations. 
In some delicate way, which we do not fully understand, the constraints 
on the wavevector summations in the vulcanization theory play a similar 
role to the field-index contractions in the Potts theory.

Having discussed the similarities of the Potts and vulcanization 
approaches, let us now catalogue the many distinctions between them:
%----------
\hfil\break\noindent
(i)~The Potts field theory has a multiplet of $n$ real fields on
$d$-dimensional space; the vulcanization field theory has a real
singlet field living on $(n+1)$-fold replicated $d$-dimensional space.
%----------
\hfil\break\noindent
(ii)~ The Potts field theory represents a setting involving a 
{\it single\/} ensemble~\cite{REF:singleENS}, the ensemble of 
percolation configurations, whereas the vulcanization field 
theory describes a physical problem in which {\it two\/} 
distinct ensembles (thermal and disorder) play essential roles.  
As such, the vulcanization field theory is capable of providing a
unified theory not only of the transition but also of the structure,
correlations and (e.g.~elastic) response of the emerging amorphous
solid state.  This is already manifested at the mean-field level,
inasmuch as the vulcanization field theory presents an order parameter
that is far richer in its physical content that the one presented by
the Potts model. 
%----------
\hfil\break\noindent
(iii)~The entire symmetry structures possessed by the percolation and 
vulcanization field theories are quite different. 
%-----
The Potts field theory has translational and rotational invariance (in 
unreplicated space), along with the discrete symmetry of $(n+1)$-fold 
permutations of the field components. 
%-----
The vulcanization field theory has the symmetries of the independent
translations and rotations of the $(n+1)$ replicas of space, along
with the discrete symmetry of $(n+1)$-fold permutations amongst the
replicas.
%----------
\hfil\break\noindent
(iv)~The nature of the spontaneous symmetry breaking at the percolation 
and vulcanization phase transitions is distinct. 
%----- 
The percolation transition (in its Potts representation) involves the
spontaneous breaking of the ($n\to 0$ limit) of a {\it discrete\/} 
$(n+1)$-fold permutation symmetry.  
%-----
By contrast, the vulcanization transition involves the spontaneous 
breaking of the ($n\to 0$ limit of the) {\it continuous\/} symmetry 
of relative translations and rotations of the $n+1$ replicas; the 
permutation symmetry remains intact in the amorphous solid state, as does 
the symmetry of common translations and rotations of replicated space.  
%-----
Thus, the vulcanization transition is associated with the appearance 
of low-energy, long-wavelength, Goldstone-type 
excitations~\cite{REF:stability}, which we expect to lead to the 
restoration of the broken continuous symmetry in and below a lower 
critical dimension of two.  By contrast, fluctuations destroy the 
percolation transition only at and below the the lower critical 
dimension of unity.  

Whilst there are these apparent distinctions between the percolation
and vulcanization approaches, especially in low dimensions, there is
also evidence in favor of some sort of sharp correspondence between
the physics of percolation and vulcanization coming from the
computation of critical exponents near the upper critical dimension.
This apparent dichotomy can, however, be reconciled if we carefully
delineate between three logically distinct physical properties
pertaining to RCMSs and other randomly constrained systems:
%--------------------------------------
\hfil\break\noindent
(i)~macroscopic network formation (by which we mean that constraints 
are present in sufficient density to connect a nonzero fraction of the
constituents into a giant random molecule); 
%--------------------------------------
\hfil\break\noindent
(ii)~random localization (by which we mean the change in thermal
motion of a nonzero fraction of the constituents from wandering
throughout the container to fluctuating only over finite distances
from their random mean positions); and 
%--------------------------------------
\hfil\break\noindent
(iii)~the acquisition of rigidity (by which we mean the emergence
of a nonzero static shear modulus).
%--------------------------------------
\hfil\break\noindent
Within mean-field theory (and hence above six spatial dimensions),
these three properties go hand in hand, emerging simultaneously at the
phase transition.  At and below six dimensions they appear to continue
to go hand in hand (although, strictly speaking, we have not yet
investigated the issue of the acquisition of rigidity beyond
mean-field theory)\thinspace\ until one reaches two dimensions where
we believe this broad picture will change (as we shall discuss
shortly). Thus, it appears that, within the limited sphere of issues
concerning amorphous solidification that percolation-based approaches
are capable of addressing, such approaches do not lead one astray.  In
other words, the superposition of thermal fluctuations on the
positions of the constituents of the macroscopic network that emerges
as the constraint density is increased towards the phase transition
does not lead to any changes in the critical exponents governing
percolation-type quantities: disorder fluctuations appear to play a
more important role than do thermal fluctuations, as far as the
percolative aspects of the critical phenomenon are concerned.

This brings up the interesting issue of the nature of the
vulcanization transition and its relationship with the percolation
transition as the dimensionality of space is reduced to the
neighborhood of two spatial dimensions, two being the lower critical
dimension of the vulcanization transition. (The ideas reported in this
paragraph result from an ongoing collaboration 
with H.~E.~Castillo~\cite{REF:withHEC}.)\thinspace\ Indeed, the case 
of two dimensions is especially fascinating in view of the fact that
there is a conventional percolation transition in two dimensions,
whereas the thermal fluctuations are expected to be sufficiently prominent 
to destablize the amorphous solid phase, in which case the macroscopic
network formation no longer occurs simultaneously with the random
localization of constituents of the network.  It is tempting to 
speculate~\cite{REF:withHEC} that in two dimensions an anomalous 
type of vulcanization transition (not accompanied by true localization) 
continues to happen simultaneously with percolation transition.  As the
constraint density is tuned from below to above criticality, the 
amorphous solidification order parameter would remain zero, 
whereas the order-parameter correlations would change from decaying 
exponentially to decaying algebraically with distance.  One might say 
that (constraint-density controlled) cluster fragmentation (rather 
than the thermal excitation of lattice defects, as in regular 
two-dimensional melting) would be mediating the melting transition.  
If this scenario should happen to be borne out then, at sufficiently 
high crosslink densities one would have a quasi-amorphous solid 
state---the random analogue of a two-dimensional
solid~\cite{REF:DRN:DandG}---exhibiting quasi-long-range positional
order but of a random rather than regular type.  By implementing 
these ideas via an effective field theory that describes low-energy 
excitations of the amorphous solid state, we hope
to construct a picture of the vulcanization transition and the
emergent rigid state in and near two spatial dimensions.  Such a
development would complement the approach to the vulcanization
transition, presented here, which is based on expanding about the 
upper critical dimension, e.g., by providing access to critical 
exponents via an expansion about two rather than six dimensions.
%------------------------------------------------------------------
%------------------------------------------------------------------
\noindent
\section*{Acknowledgments}
\label{SEC:Acknowledgments}
%------------------------------------------------------------------
It is a pleasure to thank Karin Dahmen, Bertrand Fourcade, Eduardo 
Fradkin, Sharon Glotzer, Avi Halperin, Jos\'e Mar\'{\i}a Rom\'an, 
Michael Stone, Clare Yu, and especially Horacio Castillo for helpful 
discussions.  This work was supported by 
the U.S.~National Science Foundation through grants 
DMR99-75187 (WP, PMG) and 
NSF-DMR91-20000 (WP), 
and by the Campus Research Board of the University of Illinois. 
%------------------------------------------------------------------
%------------------------------------------------------------------
\appendix 
%------------------------------------------------------------------
%------------------------------------------------------------------
%------------------------------------------------------------------
\section{Inverse susceptibility and Ginzburg criterion}
\label{SEC:GC_diagrams}
%------------------------------------------------------------------
In order to calculate the one-loop correction to the 2-point vertex
function ${\Gamma}^{(2)}(\hat{k})$, we first calculate the self-energy
$\Sigma_n(\hat{k})$ (i.e.~the sum of all two-point
one-particle-irreducible amputated diagrams), in terms of which
${\Gamma}^{(2)}(\hat{k}) \equiv G_0(\hat k)^{-1}
-\Sigma_n(\hat{k}){\big\vert}_{n \to 0}$. To one-loop order,
$\Sigma_n(\hat{k})$ is given by the amputated diagram shown in
Fig.~\ref{FIG:ginzburg},
\begin{equation}
\Sigma_n(\hat{k}) = 
18\inte^{2}
\sum_{\hat{k}_1\in\HRS\atop(\hat{k}-\hat{k}_1\in\HRS)} 
	G_0(\hat{k}_1)\, 
	G_0(\hat{k}_1-\hat{k}).
\label{EQ:SelfEnergy}
\end{equation}
Let us emphasize the meaning of the notation: one is directed to 
sum over all replicated wavevectors 
$\hat{k}_1\in\HRS$ subject to the constraint that 
$\hat{k}-\hat{k}_1\in\HRS$; one should also bear in mind the 
fact that the external wavevector $\hat{k}$ lies in the \HRS.
This constrained summation can be expressed in terms of several
unconstrained summations (for cases in which $\hat{k}$ has nonzero 
entries in at least three replicas, i.e., lies in the $3^{+}$RS) as
\begin{equation}
\sum_{\hat{k}_1\in\HRS\atop(\hat{k}-\hat{k}_1\in\HRS)}O(\hat{k}_1)=
   \sum_{\hat{k}_1}O(\hat{k}_1)
  -\sumr\sum_{{\bf p}}O(\hat{k}_1)
  	\Big\vert_{\hat{k}_1={\bf p}\rbv^{\alpha}}
  +nO(\hat{k}_1)\Big\vert_{\hat{k}_1=\hat{0}}
  -\sumr\sum_{{\bf p}}O(\hat{k}_1)
  	\Big\vert_{\hat{k}_1={\bf p}\rbv^{\alpha}+\hat{k}} 
  +nO(\hat{k}_1)\Big\vert_{\hat{k}_1=\hat{k}}\,\,\,,
\label{EQ:RG_Cons_Sum_3+}
\end{equation} 
for any $O(\hat{k}_1)$.  
Here, $\{\rbv^{\alpha}\}_{\alpha=0}^{n}$ is the collection of unit
vectors in replicated space, so that, e.g., a generic vector $\hat{p}$
can be expressed as $\sum_{\alpha=0}^{n}{\bf
p}^{\alpha}\rbv^{\alpha}$.  When $\hat{k}$ belongs to the 2RS
[e.g.~$\hat{k}=({\bf l}^{1},{\bf l}^{2},{\bf 0},\cdots,{\bf 0})$]
there is a slight modification of Eq.~(\ref{EQ:RG_Cons_Sum_3+}) 
and, instead, we have
\begin{eqnarray}
\sum_{\hat{k}_1\in\HRS\atop(\hat{k}-\hat{k}_1\in\HRS)}O(\hat{k}_1)
&=&
   \sum_{\hat{k}_1}O(\hat{k}_1)
  -\sumr\sum_{{\bf p}}O(\hat{k}_1)
  	\Big\vert_{\hat{k}_1={\bf p}\rbv^{\alpha}}
  +nO(\hat{k}_1)\Big\vert_{\hat{k}_1=\hat{0}}
  -\sumr\sum_{{\bf p}}O(\hat{k}_1)
  	\Big\vert_{\hat{k}_1={\bf p}\rbv^{\alpha}+\hat{k}} 
\nonumber
\\
&&\qquad\qquad\qquad\quad
  +\,nO(\hat{k}_1)\Big\vert_{\hat{k}_1=\hat{k}}
  +O(\hat{k}_1)\Big\vert_{\hat{k}_1={\bf l}^{1}\rbv^{1}} 
  +O(\hat{k}_1)\Big\vert_{\hat{k}_1={\bf l}^{2}\rbv^{2}}\,\,. 
\label{EQ:RG_Cons_Sum_2}
\end{eqnarray} 
For the moment, let us focus on the case of $\hat{k}\in3^{+}$RS.  
By making use of Eq.~(\ref{EQ:RG_Cons_Sum_3+}), and subsequently 
transforming each unconstrained summation into an 
integral, we obtain
\begin{equation}
\Sigma_n(\hat{k})=
18\inte^{2}
\Big(
V^{n+1}\int d^{(n+1)d}k_1\, 
G_0(\hat{k}_1)\, 
G_0(\hat{k}_1-\hat{k}) 
%	\nonumber \\&&\qquad\qquad\qquad
-2\sumr V\int d^d p\, 
G_0({\bf p}\rbv^\alpha)\, 
G_0({\bf p}\rbv^\alpha-\hat{k})
+ 2n\,G_0(\hat 0)\,G_0(\hat{k})
\Big). 
\end{equation}
The limit of the validity of the Landau theory (i.e.~the tree-level 
approximation) can be ascertained by enquiring when the loop corrections 
to the inverse susceptibility become comparable its tree-level value. 
Thus we take the long-wavelength limit of the 
correction~(\ref{EQ:SelfEnergy}) via a sequence of wavevectors 
$\hat k$ lying in the \HRS, obtaining
\begin{equation}
\Sigma_n(\hat{k})\big\vert_{\hat{k}\rightarrow \hzero} = 
18\inte^{2}
\Big(V^{n+1}\int d^{(n+1)d} k_1\, 
{G_0}(\hat{k}_1)^{2}
-2(n+1)V\int d^d p\,{G_0}({\bf p})^2
+2n{G_0}(\hat 0)^2
\Big). 
\end{equation}
At this stage, the $n\rightarrow 0$ limit may be taken [the reason for 
this is discussed in Sec.~\ref{SEC:LandauWilson}, shortly after 
Eq.~(\ref{EQ:LG_longwave})].  In addition, the integral over the 
$(n+1)$-fold replicated space goes smoothly into an integral over the 
ordinary (i.e.~unreplicated) space.  Thus, we arrive at
\begin{equation}
\Sigma(\hat{k})\big\vert_{\hat{k} \rightarrow  \hzero}
\equiv \lim_{n \rightarrow 0} \Sigma_n(\hat{k})
	\big\vert_{\hat{k} \rightarrow  \hzero}=
18 \inte^{2}
\Big(
V\int d^d p\,{G_0}({\bf p})^2 
-2V\int d^d p\, 
{G_0}({\bf p})^2
\Big). 
\end{equation}
From this expression, we see an example of what turns out to be a 
typical effect of the exclusion of the \WRS, viz., that it reverses 
the sign relative to the unconstrained version.  By collecting this 
loop correction together with the tree-level inverse susceptibility, 
we arrive at the result that we shall use to establish the Ginzburg 
criterion: 
\begin{equation}
(N\chi)^{-1}\equiv
N^{-1}{\Gamma}^{(2)}(\hat{k})\big\vert_{\hat{k} \rightarrow \hat 0}=
{G_0}(\hzero)^{-1}
-N^{-1}\Sigma(\hat{k})\big\vert_{\hat{k} \rightarrow  \hzero}=
-2a\deltat
+18\inte^2 {V\over{N}} \int  
\frac{d^d p}{(-2a\deltat +b p^2)^2}\,\,. 
\end{equation}

We mention, in passing, that when $\hat{k}$ lies in the 2RS, we need to 
use Eq.~(\ref{EQ:RG_Cons_Sum_2}) instead of Eq.~(\ref{EQ:RG_Cons_Sum_3+}) 
in evaluating the constrained summation.  The resulting two extra terms 
in $\chi^{-1}$ turn out to be nonextensive and non-divergent at the 
transition, and thus do not change the result for the Ginzburg criterion.  
(The appearance of non-extensive terms may seem strange, but also occurs 
in the semi-microscopic theory of RCMSs, where the free energy for the
saddle point value of the order parameter has a non-extensive part;
for a discussion of this issue see Sec.~2.6 of Ref.~\cite{cross}.)
%------------------------------------------------------------------
%------------------------------------------------------------------
\section{Subleading elements: Additional 
semi-microscopically generated fields and vertices}
\label{SEC:Subleading}
%------------------------------------------------------------------
The inspiration for the minimal model, Eq.~(\ref{EQ:LG_longwave}),
discussed in Sec.~\ref{SEC:LandauWilson}, comes from experience with
detailed statistical-mechanical investigations of various
semi-microscopic models of RCMSs and related
systems~\cite{epl,cross,REF:endlink,manifolds}.  The field theories
obtained in these investigations contain additional fields and vertices 
beyond those featuring in the minimal model.  Amongst them are: the 
\WRS\ field [variously denoted as $\Omega({\bf k}\rbv^\alpha)$ or 
$\Omega^\alpha({\bf k})$], which describes density fluctuations; 
various vertices that couple the \WRS\ field to itself and to the 
\HRS\ field; and quartic or higher-order \HRS\ vertices.  In
the present section we discuss the role of these additional fields
and vertices.  We shall confine our attention to effects that show up
at the one-loop level.  To avoid confusion we shall, in this section, 
denote the bare \HRS\ and \WRS\ correlators respectively by 
$G_0^{\rm HRS}$ and $G_0^{\rm 1RS}$.
%--------------------------
\begin{figure}
\vskip0.50cm
 \epsfxsize=3.0truein 
  \centerline{\epsfbox{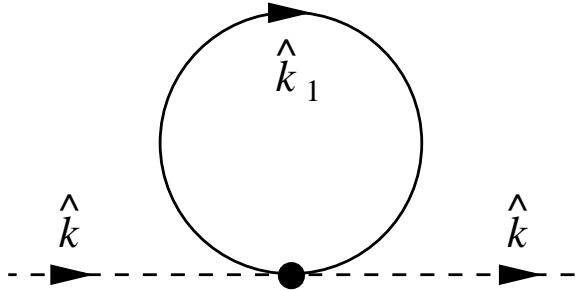}} 
% \centerline{\epsfbox{fig4.ps}} 
 \vskip0.50cm
\caption{Example of a one-loop correction to the self-energy 
	 due to a vertex omitted from the minimal model.
\label{FIG:ginzburg_sub}}
\end{figure}
%--------------------------
%------------------------------------------------------------------
\subsection{Subleading influences on the higher replica sector}
\label{SEC:SubHigher}
%------------------------------------------------------------------
We begin by considering the possible corrections to the \HRS\
self-energy $\Sigma_{n}(\hat{k})\vert_{{\hat k}\to{\hat 0}}$ arising
from the additional fields and vertices.  At the one-loop level, the
only contributions arising from an omitted vertex are those associated
with the quartic vertex, for which there are two situations to
consider, depending on whether the loop wavevector lies in the
\WRS\ or the \HRS.  Figure~\ref{FIG:ginzburg_sub} shows the relevant
diagram.  

Let us first look at the contribution of this diagram when the loop 
wavevector lies in the \HRS.  In this case, evaluating the diagram 
involves the constrained summation:
\begin{eqnarray}
\sum_{ \hat {k} \in {\HRS}} G_0^{\rm HRS}(\hat {k})
&=& 
\sum_{\hat{k}} G_0^{\rm HRS}(\hat {k})- 
\sumr \sum_{{\bf k}}G_0^{\rm HRS}({\bf k}\rbv^{\alpha})+ 
n G_0^{\rm HRS}(\hat {0}) 
\nonumber \\
&=&
\sum_{\hat{k}} G_0^{\rm HRS}(\hat {k}) - 
(n+1) \sum_{{\bf k}} G_0^{\rm HRS}( {\bf k}) + nG_0^{\rm HRS}(\hat {0}),
\label{EQ:HS} 
\end {eqnarray}
which vanishes in the $n\rightarrow 0$ limit.

Let us now look at the contribution of this diagram when the loop
wavevector lies in the \WRS.  In this case, no critical bare 
correlators feature, so that the resulting contribution to
$\Sigma_{n}(\hat{k})\vert_{{\hat k}\to{\hat 0}}$ is finite.  There are
also contributions to $\Sigma_{n}(\hat{k})\vert_{{\hat k}\to{\hat 0}}$
arising from one-loop diagrams involving two cubic vertices, in which
either one or both loop-wavevectors lie in the \WRS.  None of these
contributions alters the Ginzburg criterion established in
Sec.~\ref{SEC:Ginzburg}.
%------------------------------------------------------------------
%------------------------------------------------------------------
\subsection{Absence of feedback of critical 
fluctuations on the density-density correlator}
\label{SEC:SubFields}
%------------------------------------------------------------------
As we have discussed in Sec.~\ref{SEC:MeanField}, the \WRS\ field
$\Omega({\bf k}\rbv^\alpha)$, which describes density fluctuations,
remains \lq\lq massive\rq\rq\ at the vulcanization transition
(i.e.~the coefficient of the term quadratic in this field remains
positive at the transition), and the corresponding bare correlator is
nonsingular at the vulcanization transition.  We now examine the effects 
of \HRS\ critical fluctuations on the correlator of the \WRS\ field.  
We approach this issue by studying those one-loop diagrams for the \WRS\
self-energy in which at least one internal wavevector lies in the
\HRS; there are three types of contribution to consider:
\hfil\break\noindent 
(i)~There is the contribution associated with the diagram shown in 
Fig.~\ref{FIG:ginzburg_sub} but with the external wavevectors now lying 
in the \WRS.  By the same reasoning that we applied to Eq.~(\ref{EQ:HS}), 
this contribution vanishes in the $n\rightarrow 0$ limit.
\hfil\break\noindent 
(ii)~There are the two contributions associated with the type of diagram 
shown in Fig.~\ref{FIG:ginzburg}.  When one of the internal wavevectors 
lies in the \WRS\ and the other lies in the \HRS, the contribution
involves a constrained summation over $\hat{k}$ with $\hat{k}\in{\HRS}$ 
and $({\bf p}\rbv^{\alpha}-\hat{k})\in\WRS$ (where $\hat k$ is the loop 
wavevector and ${\bf p}\rbv^\alpha$ is the external wavevector).  In this 
case, the constraints on the summation require that $\hat{k}\in 2$RS and 
$\hat{k}={\bf p}\rbv^\alpha+{\bf l}\rbv^\beta$, where 
$\beta\neq\alpha$ and ${\bf l}\neq{\bf 0}$.  
Then, the contribution to the \WRS\ self-energy reads
\begin{eqnarray}
\sum_{\hat{k}\in\HRS\atop({\bf p}\rbv^\alpha-\hat{k}\in\WRS)} 
	G_0^{\rm HRS}(\hat k)\,G_0^{\rm 1RS}({\bf p}\rbv^{\alpha}-\hat{k})
&=&
\sum_{\beta(\neq\alpha)}
\sum_{{\bf l} \neq {\bf 0}}
	G_0^{\rm HRS}({\bf p} \rbv^\alpha + {\bf l}\rbv^\beta)\,
	G_0^{\rm 1RS}(-{\bf l}\rbv^\beta )
\nonumber \\
&=& n \sum_{{\bf l} \neq {\bf 0}} 
	G_0^{\rm HRS}({\bf p} \rbv^\alpha + {\bf l}\rbv^\beta)\,
	G_0^{\rm 1RS}(-{\bf l}\rbv^\beta)\vert_{\beta \neq \alpha}\,\,, 
\label{EQ:HSWS}
\end{eqnarray}
which evidently vanishes in the $n\rightarrow 0$ limit.  On the other
hand, when both internal wavevectors lie in the \HRS, the contribution
involves the constrained summation over $\hat{k}$ with 
$\hat{k}\in{\HRS}$ and 
$({\bf p}\rbv^\alpha-\hat{k})\in{\HRS}$.  
In this case, the contribution to the \WRS\ self-energy reads
\begin{equation}
\sum_{\hat{k}\in\HRS\atop({\bf p}\rbv^\alpha-\hat{k}\in\HRS)} 
\!\!\!\!
	G_0^{\rm HRS}(\hat k)\,
	G_0^{\rm HRS}({\bf p}\rbv^{\alpha}- \hat{k})
=\sum_{\hat {k} \in {\HRS}}
\!\!
	G_0^{\rm HRS}(\hat k)\,
	G_0^{\rm HRS}({\bf p}\rbv^{\alpha}- \hat{k})- 
\!\!
 \sum_{\hat{k}\in\HRS\atop({\bf p}\rbv^\alpha-\hat{k}\in\WRS)}
\!\!\!\!
	G_0^{\rm HRS}(\hat k)\,
	G_0^{\rm HRS}({\bf p}\rbv^{\alpha}- \hat{k}) 
\propto n\,,
\end{equation}
which also evidently vanishes in the $n\rightarrow 0$ limit.  [In the
last step we have used Eq.~(\ref{EQ:HSWS}), as well the strategy for
handling constrained summations employed in Eq.~(\ref{EQ:HS}).]
  
We conclude that, to one-loop order, the \WRS\ self-energy does not
acquire any singular contributions due to critical fluctuations in the
\HRS.  In this sense, the two sectors are well separated in the
neighborhood of the vulcanization transition.  However, it is 
straightforward to show~\cite{REF:FourField} that there are \WRS\  
correlators, such as those involving four \WRS\ fields but only two 
replica indices, which do become long ranged at the vulcanization 
transition and are thus capable of signaling the transition. 
%------------------------------------------------------------------
%------------------------------------------------------------------
\section{Derivation of flow equations 
within the epsilon expansion}
\label{SEC:RG_floweq}
%------------------------------------------------------------------
%------------------------------------------------------------------
\subsection{Implementation of the momentum-shell RG}
\label{SEC:RG_procedure}
%------------------------------------------------------------------
The first step in the momentum-shell RG approach that we are adopting
is to integrate out Fourier components of the field $\ofield{k}$
having wavevectors in the shell 
$\Lambda/b < \vert \hat k \vert < \Lambda$.  
To do this, we define $\Omega^{<}$ and $\Omega^{>}$, respectively the 
long and short wavelength components of $\ofield{k}$, by
%------------------------------------------------------------------
\begin{mathletters}
\begin{eqnarray}
\oopfield{<}{k} 
&=&
\cases{%
0,	   &for $\Lambda/b < \vert \hat k \vert < \Lambda$;  \cr
\ofield{k},&for \phantom{$/b\,$}$0 < \vert \hat k \vert < \Lambda/b$;}
	\\
\oopfield{>}{k}
&=&
\cases{%
\ofield{k},&for $\Lambda/b < \vert \hat k \vert < \Lambda$;  \cr
0,	   &for	\phantom{$/b\,$}$0 < \vert \hat k \vert < \Lambda/b$.}
\end{eqnarray}% 
\end{mathletters}% 
%------------------------------------------------------------------
Then, by exchanging 
$\ofield{k}$ for 
$\oopfield{>}{k}$ and 
$\oopfield{<}{k}$ 
in Eq.~(\ref{EQ:RG_ft}) we can re-express the effective Hamiltonian as 
\begin{mathletters}
\begin{eqnarray}
\action\big(\{\Omega\}\big) 
&=&
\action\big(\{\Omega^{<}\}\big)
	+ N \sum_{\hat{k}\in {\HRS}} 
	\Big(-\deltat+\frac{1}{2}|\hat{k}|^2\Big)
		\big\vert\oopfield{>}{k}\big\vert^{2}  
	- V\big(\{\Omega\}\big), 
\\
V\big(\{\Omega\}\big) 
&\equiv&
N\inte\,
\sum_{\hat{k}_1,\hat{k}_2,\hat{k}_3\in {\HRS}}\,
\delta_{{\hat{k}_1}+{\hat{k}_2}+{\hat{k}_3},{\hat{0}}}
\Big(
\ooofield{>}{k}{1}\,
\ooofield{>}{k}{2}\,
\ooofield{>}{k}{3} 
\nonumber\\
&&\qquad 
+\,3\ooofield{<}{k}{1}\,
    \ooofield{>}{k}{2}\,
    \ooofield{>}{k}{3} 
+  3\ooofield{<}{k}{1}\,
    \ooofield{<}{k}{2}\,
    \ooofield{>}{k}{3}
\Big)\,.
\end{eqnarray}%
\end{mathletters}% 
Now, focusing on the partition function, we integrate out the 
aforementioned short-wavelength field components $\Omega^{>}$
in the context of a cumulant expansion in $V$.  
Thus, Eq.~(\ref{EQ:Partition}) 
becomes
\begin{mathletters}
\begin{eqnarray}
[Z^{n}]
&&\,\propto
\int\dmhrs^{<}\,\exp\big(-\action^{<,{\rm eff}}),
\\
\action^{<,{\rm eff}}\big(\{\Omega^{<}\}\big)
&&\,\equiv
\action 
\big(\{\Omega^{<}\}\big)
-\ln\big\langle \exp V \big\rangle_{>}\,,
\\
\ln\big\langle \exp V \big\rangle_{>}
&&\,\equiv
\ln
\left\{
{\displaystyle
\int\dmhrs^{>}\,
\exp\Big(-N \sum\nolimits_{\hat{k}\in{\HRS}} 
	\big(-\deltat+\frac{1}{2}|\hat{k}|^2\big)
	\big\vert \oopfield{>}{k}\big\vert^{2} 
    \Big)\,\exp V
	\over{
\displaystyle
\int\dmhrs^{>}\,
\exp\Big(-N \sum\nolimits_{\hat{k}\in{\HRS}} 
	\big(-\deltat+\frac{1}{2}|\hat{k}|^2\big)
	\big\vert \oopfield{>}{k}\big\vert^{2} 
    \Big)}}
\right\}
\nonumber
\\
&&\approx
	\big\langle V\big\rangle_{>}+ 
	\frac{1}{2!}\Big(
		 \big\langle V^2\big\rangle_{>} 
		-\big\langle V  \big\rangle_{>}^{2}
		    \Big)+
	\frac{1}{3!}\Big(
		  \big\langle V^3\big\rangle_{>} 
		-3\big\langle V  \big\rangle_{>}
		  \big\langle V^2\big\rangle_{>}
		+2\big\langle V  \big\rangle_{>}^{3}
		    \Big) 
		+ {\cal O}(V^4).
\end{eqnarray}
\end{mathletters}% 
Note that we have not explicitly given the factor associated with Gaussian 
fluctuations in the wavevector shell because it is nonsingular and, 
therefore, does not contribute to the quantities that we are focusing on, 
viz., the RG flow equations.

Next, we calculate $\action^{<,{\rm eff}}$ to the one-loop level by 
computing the cumulant expansion to ${\cal O}(V^3)$ and discarding 
operators that are irrelevant in the vicinity of $d=6$. 
This amounts to retaining only terms of the form of those present 
in the original minimal model, and thus we are in a position to 
begin the task of recasting the resulting theory in its original form. 
The terms that must be considered correspond to the diagrams shown 
in Fig.~\ref{FIG:renorm}, and are computed in Sec.~\ref{SEC:RG_diagrams}.
When included, they produce the following intermediate form for the 
effective coarse-grained Hamiltonian: 
\begin{equation}
\action^{<, {\rm eff}} =  
\action^{<}- 
\sum_{\hat{k} \in {\HRS}}
f_2(\hat{k})\,
\big\vert\oopfield{<}{k}\big\vert^{2}-
\sum_{\hat{k}_1,\hat{k}_2,\hat{k}_3\in {\HRS}} 
f_3(\hat{k}_1,\hat{k}_2,\hat{k}_3)\, 
\ooofield{<}{k}{1}\,
\ooofield{<}{k}{2}\, 
\ooofield{<}{k}{3}\,
\delta_{{\hat{k}_1}+{\hat{k}_2}+{\hat{k}_3}, {\hat{0}}}\,,
\label{EQ:Change_Act}
\end{equation}
where the functions $f_2$ and $f_3$ can be found in
Sec.~\ref{SEC:RG_diagrams}.  In fact, only their long wavelength parts
are needed, i.e., we shall only need the constants $f_2^{(0)}$,
$f_2^{(1)}$ and $f_3^{(0)}$ in the Taylor expansions 
\begin{mathletters}
\begin{eqnarray}
f_2(\hat{k}) &\approx &  f_2^{(0)} + 
	\frac {1}{2} f_2^{(1)} |\hat{k}|^2  + {\cal O}( \hat k^4), 
\\
f_3(\hat{k}_1,\hat{k}_2,\hat{k}_3) & \approx & f_3^{(0)} + 
	{\cal O}(\hat k_1^2,\hat k_2^2, \hat k_3^2, \hat k_1 \cdot \hat k_2,
		\hat k_1 \cdot \hat k_3,\hat k_2 \cdot \hat k_3).
\end{eqnarray}
\end{mathletters}
The next step is to rescale 
$\Omega^{<}$ and $\hat{k}$ via
\begin{mathletters}
\begin{eqnarray}
\oopfield{<}{k}&=& z\, \Omega^{\prime}({\hat k}^\prime), 
\\
{\hat{k}}^{\prime}&=& b\, {\hat{k}}.
\end{eqnarray}
\end{mathletters}
The recasting of the theory in its original form also involves the 
restoration of the wavevector lattice, as discussed in 
Sec.~\ref{SEC:RG_philosophy}.  Having made this restoration, we 
arrive at recursion relations for $\deltat$ and $\inte$, along 
with the condition that the  coefficient of the gradient term 
be restored to its original value: 
\begin{mathletters}
\begin{eqnarray}
\deltat^{\prime} 
&=&
\big( \deltat + f_2^{(0)}/N \big)z^2 b^{-(n+1)d}, 
\\
\inte^{\prime} &=& \big(\inte+ f_3^{(0)}/N \big)z^3b^{-2(n+1)d}, 
\\
1 &=& \big(1-f_2^{(1)}/N \big)z^2 b^{-(n+1)d-2}.
\end{eqnarray}
\end{mathletters}
The computation of the coefficients in the recursion relations
simplifies under the convenient choice of $b=1+x$ with $x$ positive
and very small, because it allows the approximation of the shell
integrals by the product of end-point values of the integrands and the
shell volumes.  Thus, we arrive at the differential RG recursion
relations (i.e.~flow equations)~given in the main text in
Eqs.~(\ref{EQ:ctmflowdel}) and (\ref{EQ:ctmflowint}), along with the
coefficients~(\ref{EQ:FlowCoe}).
%------------------------------------------------------------------
%------------------------------------------------------------------
\subsection{Evaluation of two diagrams}
\label{SEC:RG_diagrams}
%------------------------------------------------------------------
The renormalizations of $\deltat$ and the gradient term acquire a 
nontrivial contribution associated with diagram~(a) of 
Fig.~\ref{FIG:renorm}, which determines $f_2(\hat{k})$ in 
Eq.~(\ref{EQ:Change_Act}).  Thus, including the symmetry factor 
of the diagram, we need to evaluate
\begin{equation}
f_2(\hat{k}) =  9\inte^2  
\sum_{\hat{k}_1 \in {\HRS}
	\atop (\hat{k}_1-\hat{k}\in {\HRS})} 
	G_0(\hat{k_1})\, 
	G_0(\hat{k}_1-\hat{k}). 
\end{equation}
We have encountered this kind of constrained summation in
App.~\ref{SEC:GC_diagrams}, and we use the recipe given there, together
with the facts that the external wavevector satisfies 
$\vert \hat{k}\vert \in(0,\Lambda/b)$ whereas the internal wavevectors 
satisfy $ \vert\hat{k}_1 \vert \in (\Lambda/b, \Lambda)$ and 
$\vert\hat{k}_1-\hat{k}\vert \in (\Lambda/b, \Lambda)$.  
In practice, we are concerned
with the small-$\hat{k}$ behavior of $f_2(\hat{k})$, in which case the
latter two constraints are equivalent (the difference in their effects
being sub-dominant).  Thus, by invoking Eq.~(\ref{EQ:RG_Cons_Sum_3+})
we arrive at
\begin{eqnarray}
&&
f_2(\hat{k}) = 9\inte^2 
\left( 
	\frac {V^{n+1}}{{(2\pi)}^{(n+1)d}}
	\int\limits_{{\Lambda/b}<\vert\hat{k}\vert<{\Lambda}}
		d^{(n+1)d}k_1\,G_0(\hat{k_1})\, 
		               G_0(\hat{k}_1-\hat{k}) 
\right.
\nonumber
\\
&&\qquad\qquad\qquad\qquad
\left.		
	- 2 \sumr\frac{V}{{(2\pi)}^d}
	\int\limits_{{\Lambda/b}<\vert{\bf p}\vert<{\Lambda}}
		d^d p\,
			 G_0({\bf p}\rbv^{\alpha})\, 
			 G_0({\bf p}\rbv^{\alpha}-\hat k)
\right).
\label{EQ:FtwoInt}
\end{eqnarray}
Then, by expanding for small $\hat k$ and taking the 
$n\rightarrow 0$ limit, we obtain
\begin{mathletters}
\begin{eqnarray}
f_2^{(0)}
&=&  
	-\frac {9}{4} \inte^2 V 
	\frac{S_d}{(2\pi)^d} 
	\int_{\Lambda/b}^{\Lambda}  
	\frac {k^{d-1} dk}{(-\deltat + k^2/2 )^2}  
	+{\cal O}(\inte^4),
	\\
f_2^{(1)} 
&=&
	 -\frac {9}{4} \inte^2 V\frac{S_d}{(2\pi)^d} 
	\bigg( 
		-\int_{\Lambda/b}^{\Lambda}  
		\frac {k^{d-1} dk}{(-\deltat+ k^2/2 )^3} 
		+ {2\over{d}} 
		\int_{\Lambda/b}^{\Lambda} 
		\frac {k^{d+1} dk}{(-\deltat + k^2/2 )^4} 
	\bigg) + {\cal O}(\inte^4),  
\end{eqnarray}%
\end{mathletters}%
where $S_d$ is the surface area of a $d$-dimensional sphere 
of unit radius.  

The renormalization of $\inte$ acquires a 
nontrivial contribution associated with diagram~(b) of 
Fig.~\ref{FIG:renorm}, which determines 
$f_3(\hat{k}_1,\hat{k}_2,\hat{k}_3)$ in 
Eq.~(\ref{EQ:Change_Act}).  Thus, including the symmetry factor 
of the diagram, we need to evaluate
\begin{equation}
f_3(\hat{k}_1,\hat{k}_2,\hat{k}_3) = \frac{8}{3!} (3\inte)^3   
	\sum_{{{\hat {k}\in {\HRS}}
	\atop{({\hat {k}}+{\hat {k}}_2\in {\HRS})}} 
	\atop{({\hat {k}}-{\hat {k}}_1 \in {\HRS})}}
	G_0(\hat{k})\, 
	G_0(\hat{k}+\hat{k}_2)\, 
	G_0(\hat{k}-\hat{k}_1). 
\end{equation}
This constrained sum is similar to the one analyzed in the context of
Eq.~(\ref{EQ:RG_Cons_Sum_3+}), but is more lengthy, yielding
\begin{eqnarray}
f_3(\hat{k}_1,\hat{k}_2,\hat{k}_3) 
&=&
36\inte^3
	\Bigg(
	\frac{V^{n+1}}{{(2\pi)}^{(n+1)d}}
	\int\limits_{{\Lambda/b}<\vert\hat{k}\vert<{\Lambda}}
	d^{(n+1)d}k\,
	G_0(\hat{k})\,G_0(\hat{k}+\hat{k}_2)\,G_0(\hat{k}-\hat{k}_1)
\nonumber
\\
&&\qquad\qquad
	-\sumr\frac{V}{(2\pi)^d} 
	\int\limits_{{\Lambda/b}<\vert{\bf p}\vert<{\Lambda}} d^d p\, 
	G_0(\hat{k})\,G_0(\hat{k}+\hat{k}_2)\,G_0(\hat{k}-\hat{k}_1)
	\vert_{{\hat k} = {\bf p}\rbv^\alpha}  
\nonumber 
\\
&&\qquad\qquad
	-\sumr\frac{V}{(2\pi)^d} 
	\int\limits_{{\Lambda/b}<\vert{\bf p}\vert<{\Lambda}} d^d p\, 
	G_0(\hat{k})\,G_0(\hat{k}+\hat{k}_2)\,G_0(\hat{k}-\hat{k}_1)
	\vert_{{\hat k}={\bf p}\rbv^\alpha-{\hat k}_2}
\nonumber 
\\
&&\qquad\qquad
	-\sumr\frac{V}{(2\pi)^d} 
	\int\limits_{{\Lambda/b}<\vert{\bf p}\vert<{\Lambda}} d^d p\, 
	G_0(\hat{k})\,G_0(\hat{k}+\hat{k}_2)\,G_0(\hat{k}-\hat{k}_1)
	\vert_{{\hat k} = {\bf p}\rbv^\alpha+{\hat k}_1} 
	\Bigg). 
\end{eqnarray}
In fact, what we need is the $n\rightarrow 0$ limit of 
$f_3(\hat 0,\hat 0,\hat 0)$, which is given by 
\begin{equation} f_3^{(0)} = -9
\inte^3 V \frac{S_d}{(2\pi)^d}
	\int_{\Lambda/b}^{\Lambda} 
	\frac {k^{d-1} dk}{(-\deltat+ k^2/2)^3} + {\cal O}(\inte^5).
\end{equation}
It is worth emphasizing that, once again, the essential consequence of
the exclusion of the \WRS\ from the theory.  Without it, even signs of
all three coefficients, $f_2^{(0)}$, $f_2^{(1)}$ and $f_3$, would be
reversed, and the fixed-point structure of theory would be completely
changed.
%----------------------------
%----------------------------
  
%----------------------------
%----------------------------
%\end{multicols}

\begin{references}
\bibitem{REF:Measure}
The measure $\dmhrs$ indicates that we integrate $\ofield{k}$ with
$\hat k \in {\HRS}$ and with $\hat k$ restricted to a half-space via
an additional condition $\hat k \cdot \hat m > 0$ for some suitable 
unit $(n+1)d$-vector $\hat m$.  This is due to the fact that 
$Q(-\hat k)=Q(\hat k)^{\ast}$ (i.e.~its Fourier transform is real).
Similarly, $\dmwrs$ indicates that we integrate the \WRS\ version of
$\ofield{k}$ for $\hat k \in {\WRS}$ (similarly constrained by
reality to a suitable half-space).
\bibitem{REF:WasFoot}
One can derive such relationships, e.g., by comparing 
Eqs.~(4.17) and (5.15a) of Ref.~\cite{cross}.
\bibitem{REF:MicroOp}
In principle, one can construct a minimal model directly in terms of 
$Q({\hat k})$ by following the same general considerations that lead
to the minimal model for $\Omega(\hat k)$.  The minimal model for
$Q({\hat k})$ is equivalent in physical content to the minimal model 
for $\Omega(\hat k)$; the difference would only lie in the values
of the phenomenological parameters $a$, $b$ and $g$. In order to keep 
in line with the strategy used in Ref.~\cite{cross}, we choose 
$\Omega(\hat k)$ as the central quantity in our minimal model.
\bibitem{REF:ZGprior}
Some initial considerations of these issues were given in 
Ref.~\cite{REF:SGRFchapZG}.
\bibitem{REF:sector_disc}
The result for the inverse susceptibility depends on the route through
wavevector space through which the $\hat k \to \hat 0$ limit is taken.
Specifically, one gets distinct results for the inverse susceptibility
depending on whether or not the limit is taken via \HRS\ wavevectors
$\hat k$ that are in the two-replica sector or not.  However, both
routes lead to the same Ginzburg criterion.
\bibitem{REF:roleQR}
By contrast, the approach adopted in the present Paper applies to 
irreversibly bonded systems, such as vulcanized media (i.e.~media 
formed by the permanent random crosslinking of pre-existing 
macromolecules) and chemical gels (i.e.~media formed by the 
permanent random bonding of low-molecular--weight multi-functional 
units).  In each case there is quenched randomness (in the form of 
either the crosslinking or the bonding).
\end{references}
\end{document}